\documentclass[manuscript]{aastex}

\accepted{to the PASP}

\def\mbar{\ifmmode\overline{m}\else$\overline{m}$\fi}
\def\Mbar{\ifmmode\overline{M}\else$\overline{M}$\fi}
\def\mibar{\ifmmode\overline{m}_I\else$\overline{m}_I$\fi}
\def\MIbar{\ifmmode\overline{M}_I\else$\overline{M}_I$\fi}
\def\Nbar{\ifmmode\overline{N}\else$\overline{N}$\fi}

\def\deg{\ifmmode^\circ\else$^\circ$\fi}
\def\arcsec{\ifmmode^{\prime\prime}\else$^{\prime\prime}$\fi}
\def\arcmin{\ifmmode^{\prime}\else$^{\prime}$\fi}

\def\dmod{\ifmmode(m{-}M)_0\else$(m{-}M)_0$\fi}
\def\mM{\ifmmode(m{-}M)_0\else$(m{-}M)_0$\fi}
\def\vi{\ifmmode(V{-}I)\else$(V{-}I)$\fi}
\def\viz{\ifmmode(V{-}I)_0\else$(V{-}I)_0$\fi}
\def\EBV{\ifmmode E_{B-V}\else$E_{B-V}$\fi}

\shorttitle{Variability in NGC 2301}
\shortauthors{Howell et al.}
\received{25 April 2005}

\begin{document}

\title{A Search for Variable Stars and Planetary Occultations in NGC 2301 II:
Variability}

\author{
Steve B. Howell,\altaffilmark{1}
Cassandra VanOutryve\altaffilmark{2,3}
John L. Tonry,\altaffilmark{4}
Mark E. Everett,\altaffilmark{5}
\and 
Raelin Schneider,\altaffilmark{2,6}
}

\altaffiltext{1}{WIYN Observatory and NOAO, 950 N. Cherry Ave.,
Tucson, AZ  85726;\\ {howell@noao.edu}}
\altaffiltext{2}{NOAO, 950 N. Cherry Ave.,
Tucson, AZ  85726}
\altaffiltext{3}{Current address: Astronomy Department, University of
California, Berkeley, CA 94720; {cassy@berkeley.edu}}
\altaffiltext{4}{Institute for Astronomy, University of Hawaii,
2680 Woodlawn Drive, Honolulu, HI 96822; {jt@ifa.hawaii.edu}}
\altaffiltext{5}{Planetary Science Institute, Fort Lowell Rd., Tucson, AZ
85719; {everett@psi.edu}}
\altaffiltext{6}{Current address: Embry-Riddle Aeronautical University, 
Prescott, AZ 86301; {schne57b@erau.edu}}

\begin{abstract}
We performed R-band time series observations of the young, metal rich open
cluster NGC 2301 for 12 nights in Feb. 2004. B images were also obtained and
color magnitude diagrams, having limits of R=19.5 and B=21.5, were produced. Only a
small effort was made to determine cluster membership as our magnitude limits
are far deeper than previously published values. Our photometric precision, for
the brightest 5 magnitudes of sources, is 1-2 mmag. We determine that for the
$\sim$4000 stars which have time-series data, 56\% are variable
and of these, approximately 13\% are observed to exhibit periodic light curves
ranging from tens of minutes to days. 
We present some examples of the light curves
obtained and produce cuts in variability space based on parameters such as
color and amplitude. The percentage of variability is approximately equal
across all colors with the majority of variables having amplitudes of 0.15 
magnitudes or less. In general, redder stars show larger variability
amplitudes.
We find a smooth decline in the number of periodic variables toward longer period.
This decline is probably due to a transition from
intrinsic to extrinsic variability and, in part,
to our limited observing period of 12 nights.
Essentially all the A and F main sequence
stars in our sample are variable ($\sim$2 mmag and larger) and most present
complex light curves containing multiple periods suggestive of their inclusion
in the $\delta$Sct and $\gamma$Dor classes. 
A variable non-cluster member giant and two variable white
dwarf candidates are discussed. Our equational description of
variability is shown to be an excellent 
predictive tool for determining the cumulative fraction of variables that 
will be observed in a photometric survey.
Our entire dataset is available electronically.
\end{abstract}


\section{Introduction}

Open star clusters are fantastic laboratories for studies of stars and stellar
evolution as the universe places at our disposal hundreds to thousands of stars
which are formed, or are still forming, from a relatively uniform set of
initial conditions. Thus, each star cluster has its own of age, metal content,
and distance. Observational study is therefore
able to place the stars within a cluster into a single context, form uniform
H-R diagrams (color magnitude diagrams), and use these to understand one piece
of the total history of star formation within the cluster. Items
such as the dynamical evolution of cluster stars into field stars, the initial
and evolved mass functions, and (with fainter magnitude limits) 
a number of science issues related to low-mass stars and brown dwarfs can be
deduced. 

Our goal in this project was to study the variability of the members of the
young star cluster NGC 2301. This cluster lies in the constellation of
Monoceros (the Unicorn) almost directly on both the Galactic equator and the
Celestial equator. NGC 2301 was first studied in detail by Grubissich \& Purgathofer (1962) 
and is now known to lie 872 pc away, be 164 million years old, and
is slightly more metal rich than the sun, [Fe/H]=+0.06. Even though this cluster is
located very near the plane of the Milky Way, it has a very small reddening (0.028 mag).
Detailed information for NGC 2301 can be found at the WEBDA open cluster web site
(Mermilliod 1992).

Studying variability allows one to locate both intrinsic and extrinsic
variables. Binary stars will eclipse, if they have a fortuitous inclination, or
they can interact with one another, both items being causes of extrinsic 
variability. One
hope we had, for example, was to discover transiting extra-solar planets
orbiting cluster member stars but none were unambiguously found 
(see Tonry et al., 2005; hereafter T05). Additionally,
stars show many types of intrinsic variability caused by parameters such as
pulsation, winds, rotation modulation, and flares. Of course, combinations of
variability mechanisms occur as well.

We obtained time series R band observations for 4078 stars in six fields covering
NGC 2301. 
We find that over half
of all the stars are variable, with $\sim$7\% of the complete sample being periodic. 
T05 found that the level of variability is well described by a power law with the
cumulative number of stars with amplitude $x$ or higher, falling like
$1/x$. 
Our observational techniques and data reductions are only briefly reviewed in the
next section as they have been discussed in detail in 
T05. We present color magnitude diagrams (H-R diagrams) for all our
stars as functions of B and R magnitude, and discuss the results of this study
in terms of variability, color, periodicity, and other parameter cuts in
the dataset. Our data are available electronically. 

\section{Observations}

\begin{figure}[t]
\epsscale{0.80}
\plotone{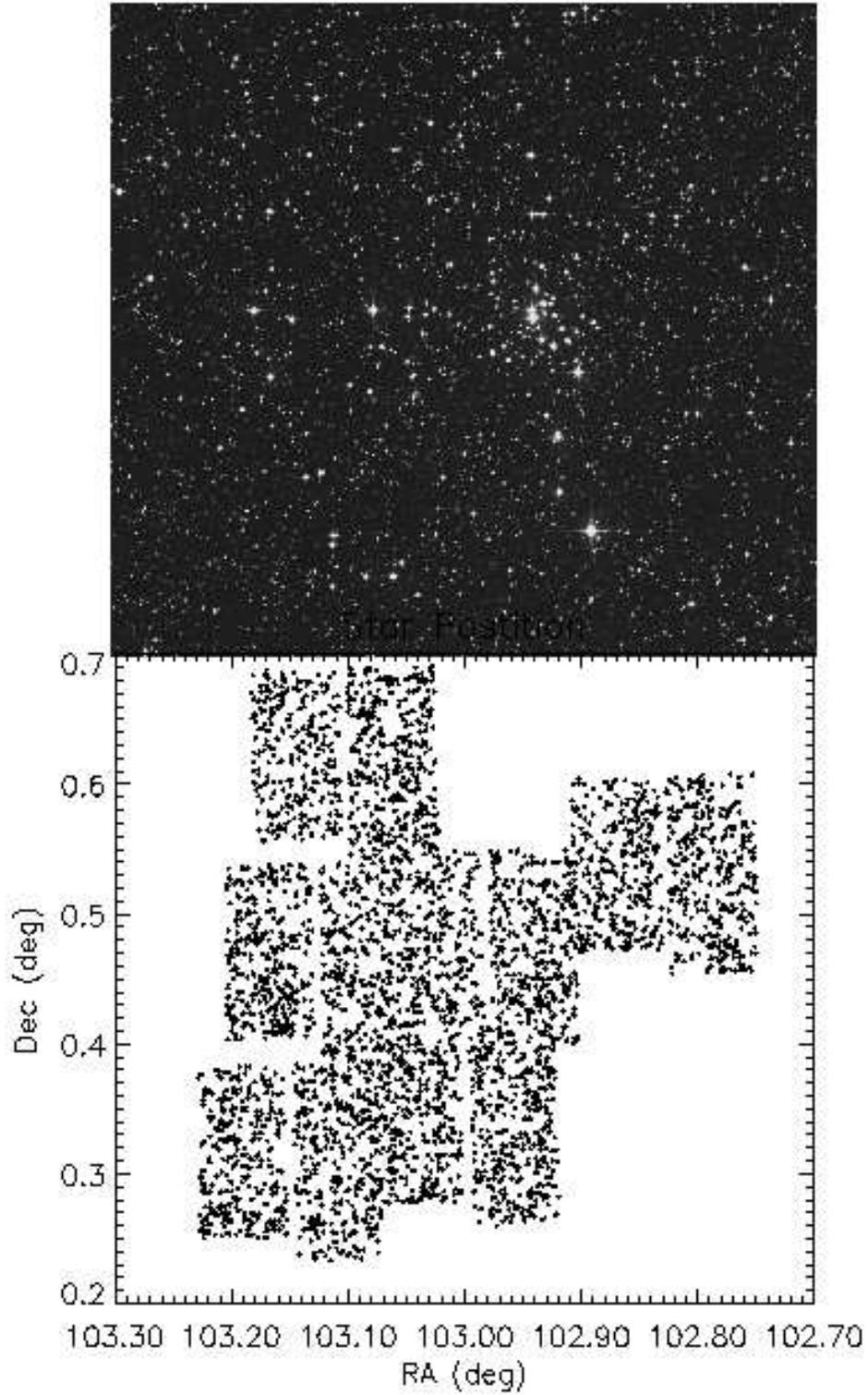}
\caption{
An image of NGC 2301 from the POSS2/red digitized sky survey (top) and the location of our six
imaged fields (bottom). The two panels are the same size and have the same field center. North is up
and East is to the left.
\label{6fields}}
\end{figure}

We obtained observations of NGC 2301 for each first half night of 12 nights
in a row (11-22 Feb. 2004) at the University of Hawaii 88" telescope atop Mauna
Kea. The very 
good quality nights of JD 2453049 and 2453052 (12 and 15 Feb. 2004) were used to
obtain B and R images of the cluster (covering a range of exposure time)
and standard star images of Rubin-152 (at various airmass values) for
photometric and astrometric calibration. 
During the remaining time, we used scripted observations to
obtain R band time series data on six adjacent (and slightly overlapping) 10 X
10 arcmin fields of view centered on NGC 2301. 
The time required to cycle
through this series of six fields was $\sim$l6 min which included 120 sec
exposures on each field, CCD readout, and offsetting to the next field. Each
half night gave us 5-6 hours on target yielding an average of 17 six-field
cycles per night.
Figure~\ref{6fields} shows the distribution of these six fields with respect to
the cluster and Table 1 lists the number of stars in each of our fields for which we
have usable data. 

The observations of NGC 2301 were made using the prototype orthogonal transfer
CCD camera called OPTIC (see Tonry et al. 1997; Howell et al. 2003). Our B and R
band CCD images of the six cluster fields (2 to 300 sec) and those
of the standard field Rubin-152 were used to calibrate all our program stars
(between $10 < R < 20.5$ and $10 < B < 22$) with final B \& R magnitudes 
having internal uncertainties of less than
0.02 mag over nearly the entire range of coverage. We note here that Rubin-152 does not contain
calibration stars redder than $B-R$=1.35 and was the only standard star field we observed. 
Our intention in this study was not to obtain highly accurate real magnitudes 
but to obtain differential light curves. Thus, the accuracy of our determined 
B and R magnitudes is likely 
to be poorer than the internal error listed above. These same B and R 
CCD images, in connection with the
USNO-B catalogue, were used to provide astrometric positions for each source good to $\pm$0.1". For
all our R band time series observations, we used one of OPTIC's special features,
PSF shaped stars, to perform our observations. The details of our observing
plan and methods and PSF
shaping are presented in detail in T05 and Howell et al. (2003) respectively.

\section{Results}

\subsection{Color Magnitude Diagrams}

\begin{figure}[t]
\epsscale{1.00}
\plotone{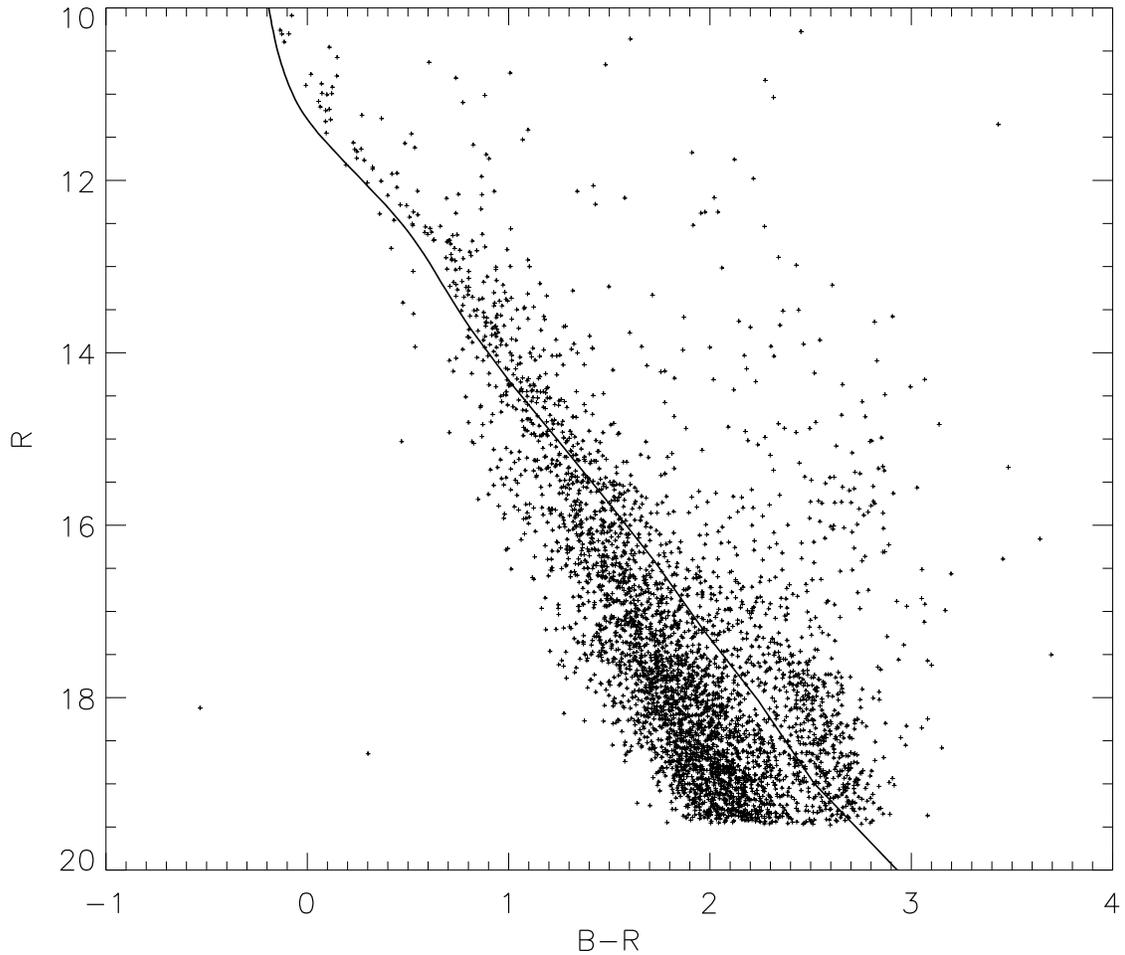}
\caption{
CMD of NGC 2301 showing R
magnitude vs. B-R color. This CMD is the sum of all six fields centered on NGC 2301 and the
solid line is the Yale isochrone as described in the text. 
The turnoff magnitude for this cluster (164 million years
old) is near R=9.5 (M$_V$=0; spectral type $\sim $B9V), just brighter than our photometry limit. Two
sources lie roughly near the expected position for cluster member white dwarfs (see text).
\label{cmd1}}
\end{figure}

\begin{figure}[t]
\epsscale{1.00}
\plotone{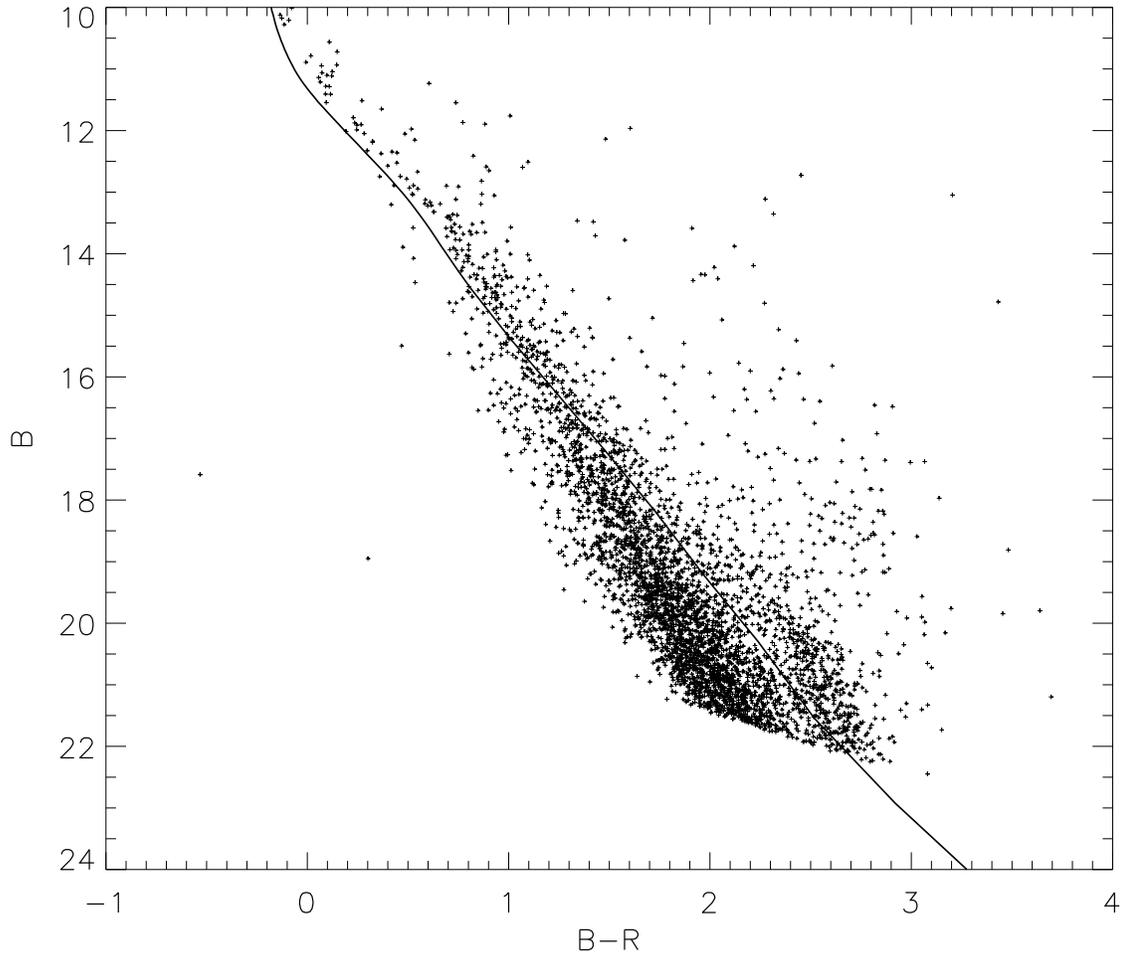}
\caption{
CMD of NGC 2301 showing B
magnitude vs. B-R color. See Figure 1 for details.
\label{cmd2}}
\end{figure}

Figures~\ref{cmd1}\&\ref{cmd2} present our complete color magnitude diagrams 
(CMD) for NGC 2301. These
figures are based on our calibrated B and R magnitudes for each star observed.
Due to our bright limit, we do not have observations of stars in NGC 2301 that lie
much above M$_V$=0 (R$\sim$10, the turnoff absolute magnitude) both on the main sequence (NCG
2301 has some blue
stragglers) and the giant branches. At our faint limit we reach to main
sequence spectral type M2V (B-R$\sim$2.8). 
The main sequence is well defined and narrow at the
bright end and begins to widen slightly around B-R=1.0 (near
F7V) and eventually (near B-R=2.0) shows a slight curve away from the previous linear 
decline (see below). The widening 
observed at fainter magnitudes is due to a combination of 
photometric uncertainty, G, K, and M star binarity, intra-cluster reddening,
and contamination by non-cluster members. All of these effects are
typical in open cluster color magnitude diagrams. 

The solid line in Figures~\ref{cmd1}\&\ref{cmd2}
is a template Yale isochrone (chosen from
the publicly available model data, 
see  http://www.astro.yale.edu/demarque/yyiso.html) to best match the 
published age and metallicity for
NGC 2301 (Z=0.02, Y=0.27, OS=0.20, l/Hp=l .743201, [Fe/H] = 0.046320,
[$\alpha$/Fe] = 0.00,
with an age of 0.1 GYr). We draw in the Yale isochrone for illustrative purposes 
only and note that it falls below the main sequence for all colors of stars 
and is known not to fit any open cluster well 
for masses below 0.6 M$_{\odot}$
(C. Deliyannis priv. comm and see Raffauf et al. 2001; Hunsch et al., 2003). 
We note that the stars on the upper main sequence
in the cluster generally follow the isochrone
shape quite well, including the kink near B-R=0.5 ($\sim$A7V) where the outer
stellar atmosphere changes from radiative to convective. At the lower end of
the main sequence, we see that the isochrone falls below the actual main sequence
around the time the stellar mass falls below 0.6 M$_{\odot}$ (late K).

The narrowness of the upper main sequence is interesting. It does not appear to
be widened (as seen for later types and attributed to binarity) nor does it
even seem to show the general spread due to small luminosity changes caused by
rotational effects (pole-on stars being slightly brighter for fast rotators).
One can imagine that a binary in this region containing a late type star may
appear single in color due to the large contrast in light with its companion.
Does this mean that all binaries in this region are composed of mismatched
luminosity pairs? What about equal mass binaries that would preserve the color of
the system? This latter idea fails as the color may remain the same but the
luminosity distribution would be different than single stars and 
would widen the main sequence distribution. Thus, the narrow upper main 
sequence is a bit of a mystery as it seems to require all the stars to be similar
with regard to binarity (or not) and type of companion.
We will discuss our
findings below for these upper main sequence stars based on their light curves,
but to end the suspense, we find no strong evidence for any binarity in these stars
and spectroscopic follow-up will likely be required to solve this mystery.

\begin{figure}[t]
\epsscale{0.80}
\plotone{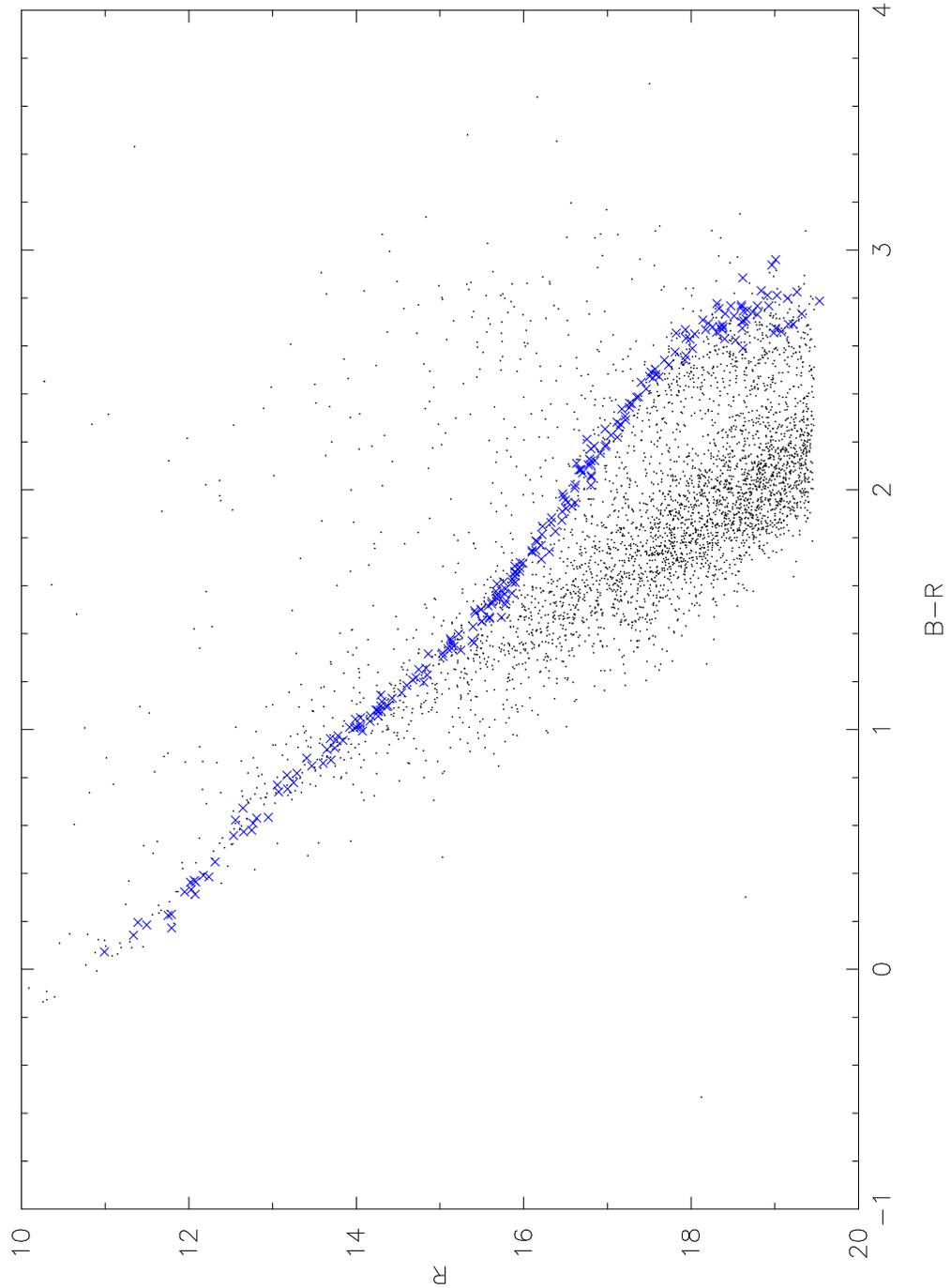}
\caption{
CMD of NCG 2301 showing our sources (black dots) and the main sequence distribution of the similar age
open cluster M34 (blue crosses). Note how the lower main sequence in both clusters curves up and to the
right compared with the linear trend of the Yale isochrone prediction (see previous figures). The ``wedge" of 
likely non-cluster members is obvious below and to the left of the main sequence. See text for details.
\label{m34}}
\end{figure}

In order to demonstrate the poor fit of the Yale isochrone at low masses as well as 
highlight the main sequence stars in NGC 2301, we over-plot our CMD of
NGC 2301 with the main sequence stellar distribution for the similar age (110-190 Myr) 
open cluster M34 (Figure~\ref{m34}) as determined by
C. Deliyannis (2005, private comm.). The M34 main sequence
was shifted in the y-direction (to accommodate the different cluster distances)
but only slightly (B-R=0.25) in the x-direction (likely due to B-R zero point
differences)
with no other corrections applied. 
The similarity of the two main sequences is quite good except
for the faintest stars where photometric uncertainties (both internal and between the two datasets)
are likely to be large. 
The purpose of Figure~\ref{m34} is to 
point out the main sequence shape in NGC 2301 and help isolate non-cluster members (see next section) as well as 
show the true lower main sequence distribution compared with that predicted by the Yale isochrone.

\subsection{Cluster Membership}

Cluster membership in NGC 2301 has been examined in some detail by Marie (1992) and other
references available at the WEBDA site (Mermilliod 1992). However, astrometric probabilities
derived from proper motion studies for cluster membership cover stars of apparent magnitude
only as faint as R$\sim$12. 
Given our bright limit, former studies are of little help is sorting out cluster membership.

An examination of the CMDs presented, especially that shown in Figure~\ref{m34}, 
allows us to make a global statement about cluster
membership which eliminates some of the stars in this study as likely non-members.
The argument goes as follows. We believe that the CMD for an open cluster is basically an H-R
diagram in which stars fall on the main sequence or in the various branches populated by
evolutionary effects. Given that few if any stars will contain less (or more) metals than the majority, 
it is rare to find stars lying below and to the left of the main sequence (unless they are fully
evolved and are white dwarfs). Stars may be reddened both by distance effects and intra-cluster
patchy reddening, but this effect moves stars to the right in a CMD. The stars near the lower
main sequence that lie to the right of it are likely reddened stars, background
stars, or photometric errors.
Given that the M34 main sequence provides a guide as to where in the CMD 
the NGC 2301 main sequence stars will fall,
we can provide a rough membership cutoff to be those stars
that lie below and to the left of the main sequence. 
Thus, we believe that most of the stars lying in the wedge shaped region 
below and to the left of our cluster main sequence are likely not to be cluster members.
The lesser number of stars lying to the right of the main sequence 
are a mix of cluster stars, reddened cluster stars, and distant 
non-members which we can not separate
cleanly. The two stars far to the left of the main sequence are 
cluster member white dwarf candidates 
and will be discussed separately.

NGC 2301 lies in the Orion arm star forming region, thus many of the stars imaged, including the
non-members, are fairly young. The typical older field stars are simply not very populous as 
Orion complex dense molecular clouds provide a curtain like backdrop 
eliminating the sampling of a large distant volume containing an
older population of background stars.
Additionally, the region surrounding NGC 2301 is known to have a
number of groups associated with it, each of which contain small collections of young stars (Mermilliod
1992). These two facts taken together limit the number of non-cluster ``wedge" stars as well as
suggest that most of the non-members are younger stars than in a typical field population.

\subsection{Variability}

We use our R band time series data of the 4078 point sources in our study to
examine variability within NGC 2301. Variability is a subjective quality that
depends on factors such as photometric precision, total time of study, and user
defined criteria which set limits of what is considered to be variable or not.
We start with the photometric precision of our data set. Examination of Figure
4 in T05 reveals that we have photometric precisions of $\la$1-2 mmag for stars
of R=11.4 to 16.5 and somewhat degraded precision after that,
reaching 0.01 mag at R=18 and 0.04 mag at R=19.5. For completeness, we have calculated
the variability level expected from atmospheric scintillation alone (see Young 1967) in order to rule
it out as a significant contributor to our highest precision (1-2 mmag) photometric results.
For our telescope, observing site, time sampling, and observing conditions, we expect scintillation to 
cause a change in the stellar intensity of 0.12 mmag, only 1/10th the level of our best precision.

The global light curve statistics shown in 
T05 (their Fig. 4) represent the
standard deviation of the entire 12 night light curve for each star. However, our data reduction
procedure (see Everett \& Howell 2001) provides a magnitude uncertainty for each datum in
each light curve\footnote{In fact we calculate two magnitude errors for each datum; one is based on 
a model of the telescope and detector and the other is the observed uncertainty based on 
the CCD S/N equation.
The observed uncertainties per datum are always equal to or slightly worse than the ideal model.}
and these detailed values must be used as
well to determine if any given point, or the entire light curve in general, is variable.

As discussed in T05, we used PSF-shaping techniques for our time series observations and
formed each star into a 30 X 30 pixel mesa. Any object that was located on a CCD edge, bad column, or
known bad pixel was eliminated from the sample up front. All remaining 4078 stars were imaged on good
CCD regions with no known defects or other issues. 
After the light curves were examined in detail, seven stars in region E (out of
678) were seen to show
a random 0.02 or less magnitude bump (drop) in their brightness for about 10-20\% of
their light curve points. The cause of this was traced to the guide star we used for region E. It was
discovered to be a very close pair of stars ($\sim$0.4" apart) which only became separated during a few 
times (about 10 frames) 
of very good seeing. Only seven stars in region E appear to be Affected as they must have been shifted onto 
pixels which did not flat field equally with the rest. 
A linear correction to the offending differential
light curve points brings these light curves into proper form. 
We did not notice any similar effects 
for any other stars in region E or for any of the other regions.
These seven stars are of
little concern in the following analysis as none were used for our ensemble
stars and the small level of occasional offset was not a factor in the
variability statistics (see T05).

An examination of Figure 4 in T05 reveals that complete saturation occurs for stars of R=10 
or brighter, while
for R fainter than 11.4 the error statistics match model predictions and a normal CCD S/N 
distribution well. However, for stars not completely saturated but brighter than R=11.4, the 
variability we observe should not be considered accurate. These stars had, at times, some part of their
bumpy square tops (see Figs. 9, 13, and 14 in Howell et al., 2003) pushed into the non-linear region of the CCD when the instantaneous seeing provided a
narrower and higher PSF. If one or more of these ``good seeing" times occurred during the 
formation of a
square PSF, the total counts from that star on that frame may be slightly unreliable. Modeling this
scenario, we find that stars fainter then R$\sim$11.4 were never affected by this while 
for those brighter than
R$\sim$11.4 (but not completely saturated) the variation induced by the non-linearity 
can be as large as 0.2 to 0.4 magnitude. Our sample of 4078 stars contains only 35 stars that 
may be effected by this phenomenon.

Each of our light curves was subjected to a battery of statistical tests.
These included tests for variability (nightly and
total), the presence of transit-like events, Monte Carlo probabilities that
random noise could account for any of the variability we detected, a gross
period search, and light curve amplitude statistics. Each of these tests
properly uses the magnitude and magnitude error of every datum over the entire
light curve in the calculations. To place a star's full multi-night light curve into the variable bin,
we adopt the following general definition. Using a reduced chi-squared value
for a constant fit to the full light curve, we take a value of 
$\chi^2$/n=1.0 to 
represent a constant star 
(true in a perfect world) and a value of $\chi^2$/n=3.0 or greater to
indicate variability at a conservative level. Note this definition allows for
stars to be continually variable or sporadically variable and places no limit on
the way in which they vary. We determine the variability statistics for each
source on a nightly basis as well as for the total light curve.
We know of no case of an object that is statistically
variable in its total light curve but is constant on any single night. 
T05 gives examples of sources we consider to be constant and ones we consider 
to be variable. The constant stars can show 
residual systematic
issues, scintillation effects, or possibly real variations but not at a level
so as to be statistically acceptable using our fairly strict
variability culling process.

In order to confirm our discovered variability, 
we performed additional tests as sanity checks to the more traditional statistical
analysis. Plots were produced that divided the stars by variable/non-variable and then further divided
the variables by amplitude of variability. This was done for each of the six fields independently.
We examined these plots for any correlation we might find such as level of variability and
non-variability with x,y location on the CCD or a stars CMD location, 
closeness of neighbors, closeness to CCD edge,
or even two-dimensional correlations such as flat fielding errors might cause. We further examined 
intra-region comparisons of
variable/non-variable and cluster/non-cluster sources to determine if specific regions 
acted differently. No correlations were found for any of the above conditions.
Therefore, we have no reason to doubt the results of our statistical determinations related to
variability for any of the stars in our sample (with the possible exception of 
sources brighter than R$\sim$11.4).

Using the above definition for
variability, we find that at the $\chi^2$/n=3.0 level or greater, 56\% of all our point sources are
variable. Note, this includes both cluster and non-cluster members. 
We did find that the percentage of variability drops off as a function of distance from the cluster
center, as expected. The number of variable sources is $\sim$75\% within 
a radius of 0.1 degree of the 
cluster center and drops to below 40\% at a radius of 0.3 degree 
from the cluster center. This change is entirely
due to cluster members with the non-member wedge star contaminates 
yielding $\sim$36\% variability in
all six fields at all radii.
Our 56\% overall variability is larger than expected for a sample of 
typical field stars but
for a young open cluster is about right. We note that studies of very young open clusters 
have revealed large percentages of variable sources as well.
For the Pleiades (125 Myr old, Scholz \& Eisloffel 2004; Martin \& Rodriguez 2000), 
it is shown that
12/26 low mass M stars were observed as variable (46\%) due to rotation modulations 
and from 19 nights of observation over 
3 years, 7/12 (58\%) of suspected upper main sequence A \& F pulsators were confirmed as variable at the
few mmag level.

\begin{figure}[t]
\epsscale{0.80}
\plotone{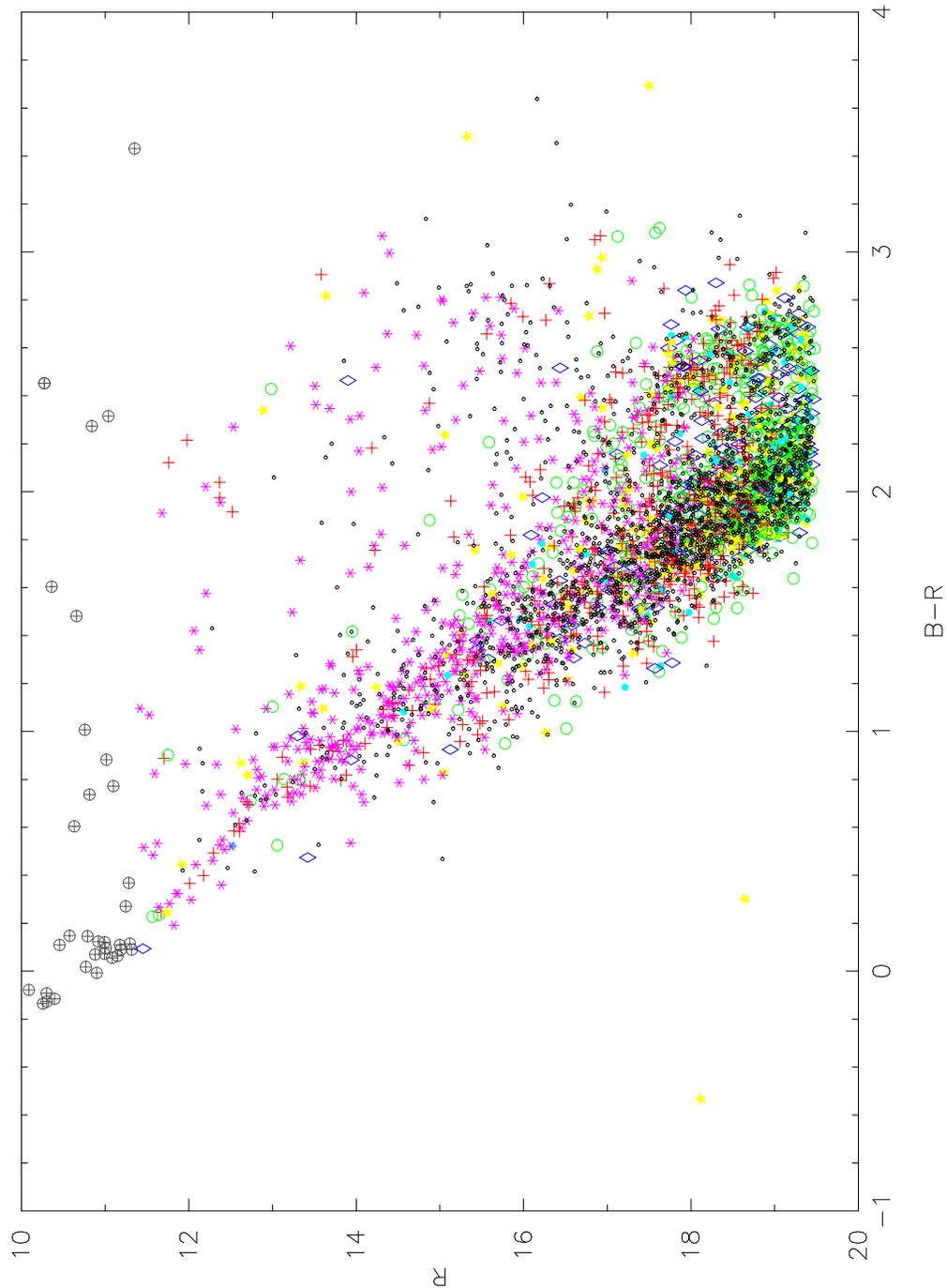}
\caption{
CMD of NGC 2301 containing the total sources from all six fields. 
The small open black circles represent constant stars while colored
symbols represent variable sources. Magenta *'s depict sources 
with $<$0.05 mag of variation, 
red plus signs are 0.05 to $<$0.1 mag,
green circles are 0.1 to $<$0.2 mag, blue diamonds are 0.2 to $<$0.3 mag,
light blue dots are 0.3 to $<$0.4 mag, and yellow stars are $\ge$0.4 mag. 
The brightest 35 stars are shown as open black circles with plus signs 
inside. These sources are 
correct in their location in the CMD but their variability statistics 
are unreliable.
\label{cmdtot}}
\end{figure}

Figure~\ref{cmdtot} 
presents our complete color magnitude diagram for NGC 2301 showing
the stars separated into variables and non-variables based on the above
definition. We have color coded the sources in Figure~\ref{cmdtot} by 
variability amplitude. Stars brighter than
R=11.4 are plotted (as their B and R magnitudes are correct) but their variability statistics are 
suspect as we have discussed above. The CMD for the open cluster NGC 2301 shows that the variability in
number, color, and amplitude is equally distributed. The only exceptions are the upper main sequence,
where we find a preference for low amplitude variations, and the fainter 
stars (R below $\sim$19) for which our statistics
disallow small amplitude detection. We have
made CMDs of variable and non-variable stars for each of our six cluster
regions and find that they show essentially no differences from the CMD shown in 
Figure~\ref{cmdtot} in their gross detail.

We present some variable star
examples below as a means of confirmation of our dataset with
respect to previous published time series results as well as illustrations of
the types of the variables within NGC 2301. Our light curves (all obtained
in R band) are presented as
{\it instrumental} R magnitude and are very high
precision differential light curves with small magnitude uncertainties per point.
We note that the conversion from instrumental R magnitude
to real R magnitude for our dataset varies by region. In region C, 
$R_{real} \sim R_{instrumental} + 1.2$ mag while all other regions have 
$R_{real} \sim R_{instrumental} + 3.0$ mag.  
Single epoch, real B and R magnitudes for all our sources (see \S2) 
are available.

\subsection{Overlap Regions}

\begin{figure}[t]
\epsscale{0.80}
\plotone{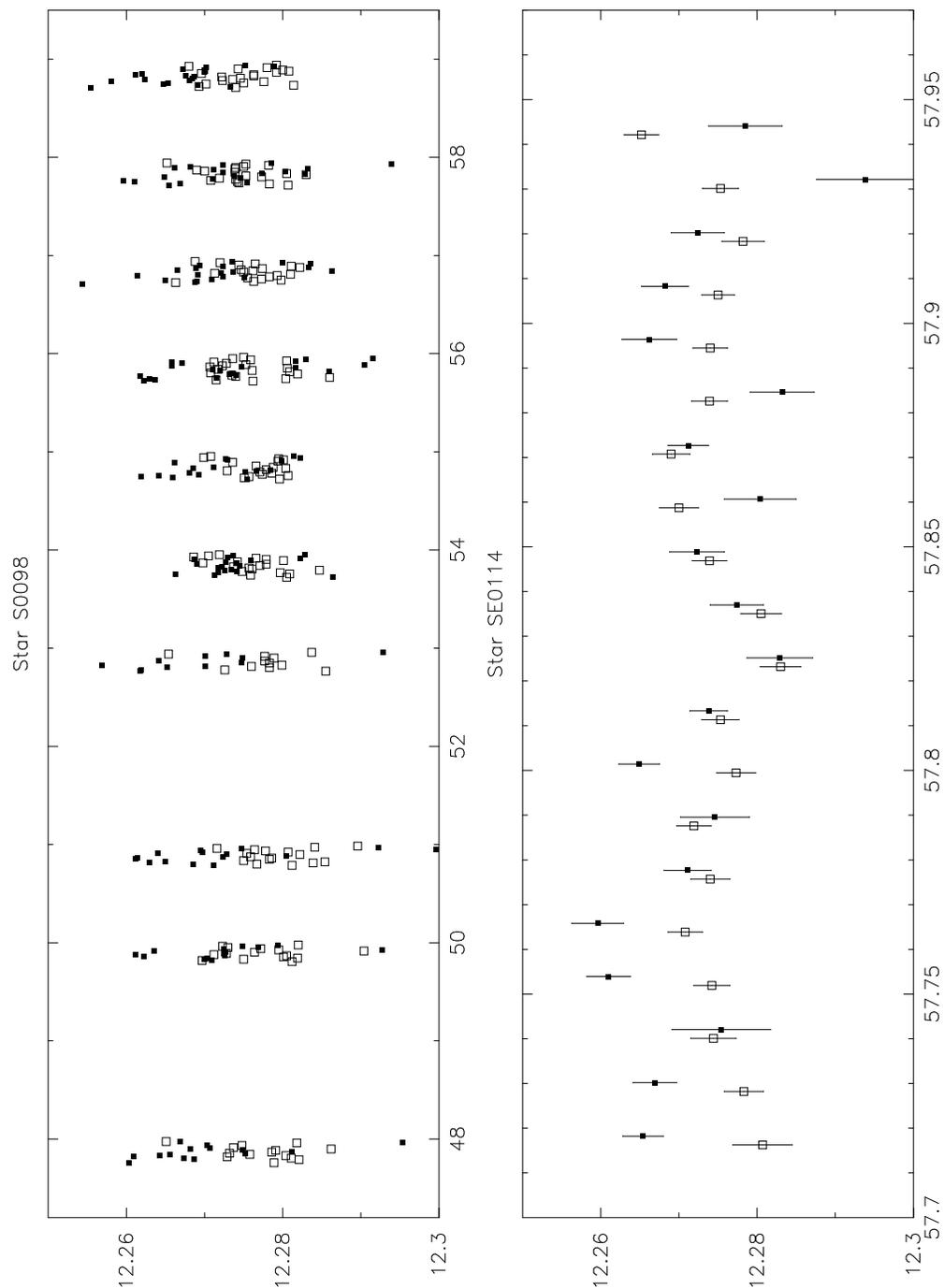}
\caption{
Example of an overlap
star. Star S0098 equals star SE0114 and we plot it here for the full light curve run (top
panel) and for a specific night (bottom). S0098 is plotted as the open squares and star SE0114
is plotted as the filled squares. We show error bars only on the bottom panel for clarity. The
y-axis is instrumental magnitude (R$\sim$16) and the x-axis is HJD-2453000 days. 
\label{olap}}
\end{figure}

An examination of Figure~\ref{6fields} shows that overlap regions occur between each of our
six imaged regions within NGC 2301. In total, 149 stars fall in the overlap
regions and we have examined these stars for both our internal accuracy of
differential light curve shape and brightness and for our formal photometric
solution to the proper R and B magnitude scale. Figure~\ref{olap} shows a typical 
example of
one of our overlap stars. We see that the light curves from the two different regions
agree
very well between the different CCD pointings even though the star was imaged on
different x,y locations and at different times. The agreement of the two light curves 
is not perfect but keep in mind that the data were not taken simultaneously (there was some minutes
between exposures) and the two data sets
are indistinguishable within their uncertainties. Examination of the B and R
magnitudes determined for the overlaps stars indicates that they agree internally to $\pm$0.02
magnitude in all cases.

\subsection{Non-Variable Sources}

\begin{figure}[t]
\epsscale{0.80}
\plotone{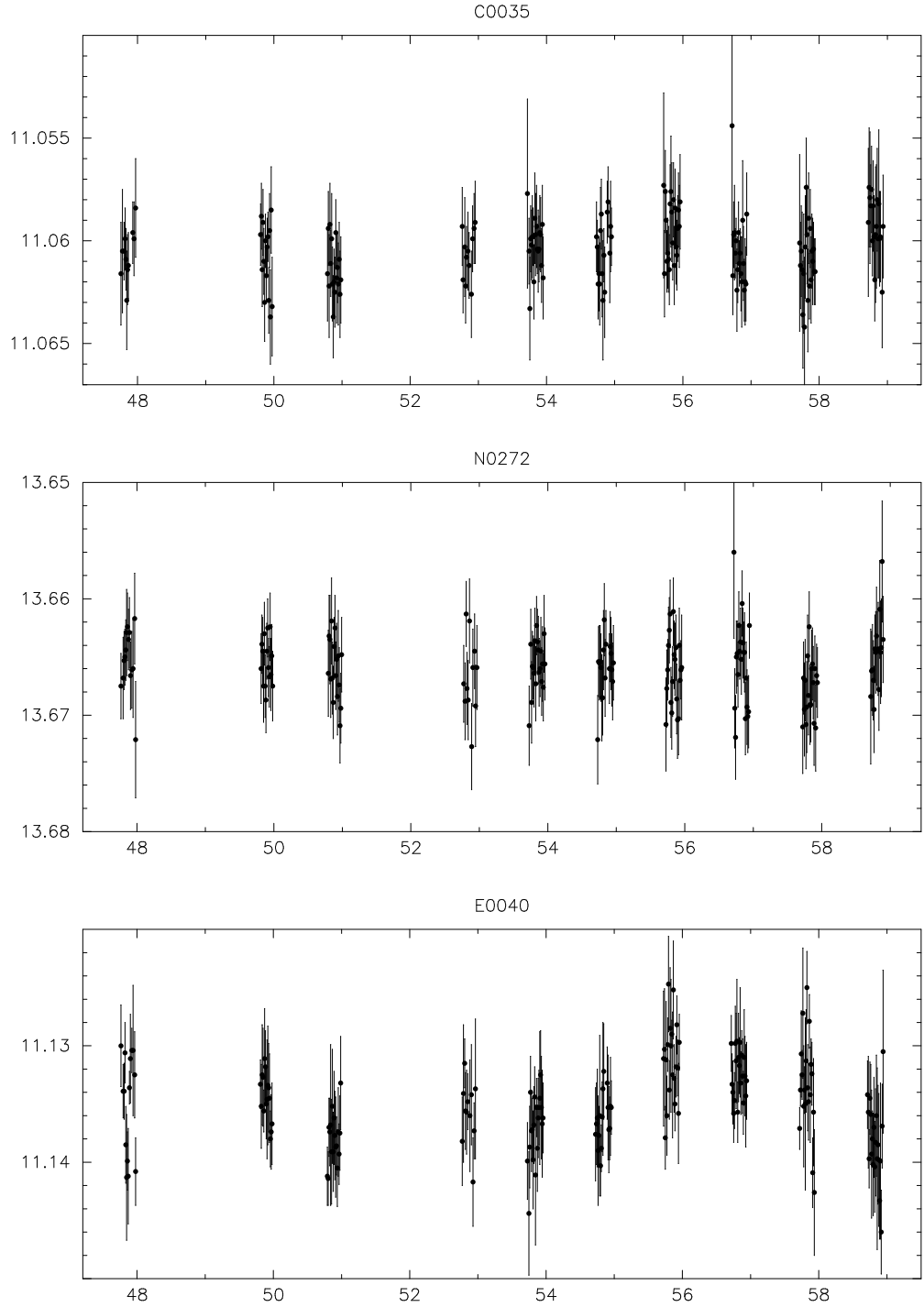}
\caption{
Three examples of
constant stars in NGC 2301. The stars shown span a range in color and variability statistic and
illustrate what we consider as constant based on our $\chi ^2$ statistic (see Table 2). The y-axis is
instrumental magnitude and the x-axis is HJD-2453000 days.
\label{cnst}}
\end{figure}

Stars within our sample that show no variability are of two kinds: stars which
truly are non-variable (if such stars exist) and those for which any
variability was not detected due to the limited time we observed them or their
variability was below our level of statistical detection. Such stars are very
important to identify as they are equally useful as probes of stellar
evolution, revealing periods of time or stellar types which are constant.
Additionally, they can be used in future studies as variability reference stars
(assuming they remain constant) to test and validate photometric and
statistical analysis. Nearly half our stars (both cluster and non-cluster
members) are found by us to be non-variable. Table 2 lists some stars we
determine to be constant based on our statistical tests (described above) and
Figure~\ref{cnst} shows a few example light curves.

\subsection{Variable Sources}

Over 2000 stars in our dataset are variable in some manner and at some level.
We discuss below a few specific cases of variability in order to provide the
reader with a flavor of the data. We do not present each light curve but rather
provide a statistical discussion of the variables including stars for which a
period is determined. One of our original goals was to search for photometric transit 
events by extra-solar planets. Our light curve statistical analysis has a number of
tests for such events but found no convincing candidates. We do see
many light curves with 0.01 mag or less modulations (about the correct level for a ``hot
Jupiter crossing a solar-type star) but none of these has a simple transit shape or a
repeatable transit-like event during our limited sampling time. 
Most of the low amplitude variables are stars with
grumbly light curves that make it difficult to interpret the cause 
in a simple manner using our time-limited dataset alone.

\subsubsection{Previously Known Variable Stars in the Central Region of NGC 2301}

A previous search for variability in NGC 2301 was undertaken. As a
step to confirm our process of variability and period searches, we
compared sources in our central (C) region to those found to be variable in Kim
et al., (2001). These authors performed a CCD V band time series study of the
central region of NGC 2301 and upon examination of 923 stars between 10th and
20th magnitude, reported nine new short period variables.
Multi-color observations were also obtained, and from the light curves and
color information, Kim et al., made identifications for the new variables, most
being eclipsing binaries and $\gamma$Dor stars. Kim et al., do not provide RA and
DEC information for their variables, only a cartoon-like image with the stars
marked. We attempted to cross reference their nine stars with stars in our C
region but were only successful at uniquely matching 7 of the 9. The other two
should be present in our data, but the chart in Kim et al., was not precise
enough to allow unambiguous identification. Table 3 lists the variable type
assigned to each star by Kim et al. 

\begin{figure}[t]
\epsscale{0.80}
\plotone{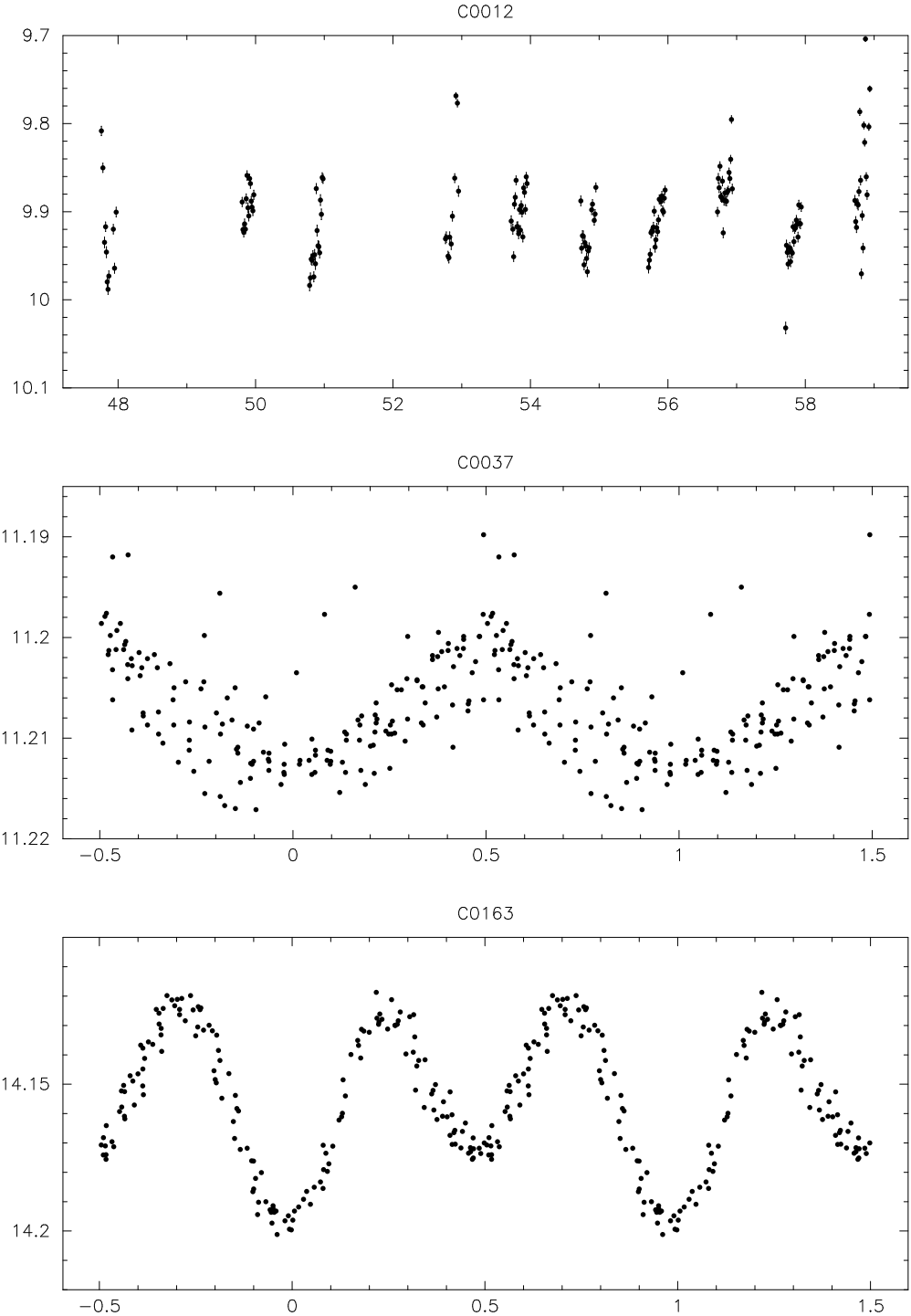}
\caption{
Three variables
identified by Kim et al. (2001) in the central region of NGC 2301. Star C0012, a suspected
slowly rotating B star does not match the period or phasing suggested in Kim et al. However,
stars C0037 (a $\gamma$ Dor star) and C0163 (an Eclipsing Binary) phase well with the
period determined in Kim et al. (see Table 3). The y-axis is instrumental magnitude and the
x-axis is HJD-2453000 days in the top panel and phase in the lower two panels. 
\label{kim}}
\end{figure}

We used periods determined in Kim et
al. to phase our matched light curves in order to examine the longevity of the periods,
assumed to be constant for the eclipsing binaries at least. Even though all of
the variables roughly agreed in total amplitude (they used V band and we used R
band for time series work), three of the stars did not maintain their phase
(even for periods within the listed uncertainties determined for our epoch).
This is not surprising, as two of the three are likely to be multiply periodic,
but we do not have an explanation for the lack of phase matching for V1, the
slowly rotating B star candidate (see Table 3). 
It may be that its type was misidentified by
Kim et al. Figure~\ref{kim} shows our light curve of their variable star C0012 (V1)
plus C0037 (V3, P=0.299 d) and C0163 (V8, P=0.250 d), the latter two phased on the Kim et al.,
ephemeris with an arbitrary zero point. We should note that our period
determinations for these last two stars agreed with the Kim et al. values to
0.001 days. Note that star C0037 (middle panel in Fig~\ref{kim}) shows the typical
``not quite perfect" phasing of a $\gamma$Dor star having multiple
periods simultaneously present (Henry \& Fekel, 2005; Arentoft et al., 2005).

\subsubsection{Periodic Variables}

Testing for a single robust period in an automatic way for a large number of stars
can be a complex process and one has to be wary of failure modes such as aliasing,
especially for periods near the data sampling time and the length of the light
curve. 
Additionally, a simple period finding approach can be fooled for systems
such as W UMa type binaries (eclipsing binaries) as the period found may really
be only one-half of the true period. 
Poretti \& Beltrame (2004) provide insight into this problem (and a good example
of a failure - see their Figure 2) as well as discuss
a solution for dealing with the problem, at least for multi-periodic light 
curves. 
We desired to determine if every light
curve was a candidate for periodicity and if so, what single period was present and at
what confidence level. We performed a gross single pass period search for each
light curve\footnote{Even the non-variable light curves were period searched
as a check on our statistical techniques. No non-variable source showed any
periodicity.} using the PDM technique (Stellingwerf 1978). Each light curve is
assigned a probability of periodicity at a very conservative level
(we used essentially 100\%) and if true, a single best fit period is determined.
This approach allowed all the light curves to be searched rapidly
with only minimal issue regarding poor period assignment or assignment as
non-periodic to multi-periodic stars (see below). Figure~\ref{pervar} shows 
an example of
four variable stars from our sample which were assigned a high probability of
periodicity. We have checked nearly 200 periodic stars in detail to see how well our
period fitting works and in all but 6 cases the stars appear to be truly
periodic. The exceptions appear to be quasi-periodic while containing other
variability.
A number of periodic stars, especially those with orbital periods 
near the length of the data set
or longer, were likely not
found by our analysis. These ``failures" are due in 
large part to the fact that 
there is a huge covariance between period and amplitude in a star's light curve and
if one does not have at least a proper inflection point to show a periodic structure it is
hard to properly identify the star as periodic without further information..

Using the above period search method, we find that 13\% of all our variable stars are 
periodic and Figures 10a \& 10b
present color magnitude diagrams for NGC 2301 separated by this property and
detailing both period and amplitude.
We see that the lower main sequence is well defined by periodic variables, likely to be
contributions from eclipsing binaries and rotational modulations.
The percent of cluster variables which are (singly) periodic is small compared with 
the number of cluster stars that are variable. This value (7\%/56\%) seems to be
roughly the same as a set of random field stars as best as we can tell (see Everett et al., 2002).

\subsubsection{Upper Main Sequence Stars}

The upper main sequence stars in NGC 2301 consist of spectral types late B/
early A to a B-R=1.0 or near F6V. Our findings (using only those upper main sequence stars that are
fainter than R$\sim$11.4, that is A0 to F6) were a surprise
as nearly every star is
variable. Examination of the light curves of stars in this class show complex
night to night behaviors with times of periodic looking data and times of
apparent randomness. Figure~\ref{mulper} presents two example light curves. 
A single
period search of the data sets in Figure~\ref{mulper} finds only a 70\% 
confidence of a single period
while the light curves of the Am star HD 8801 (Henry and Fekel
2005) provide a good comparison. HD 8801 is listed as the first example of a star which
pulsates with both $\delta$Sct and/or $\gamma$Dor frequencies. 
HD8801 shows variations of 5-10
mmag and has six simultaneous frequencies during the time of the Henry and
Fekel observations. Our upper main sequence variables have a few to 40
mmag amplitude variations and most lie in one or both of the instability strips
(see Fig 5 in Henry and Fekel). 
The occurrence of variable A-type stars
in open clusters was first looked at in detail by Abt (1979) and
searches for such variables, especially pulsating stars, 
in open clusters is reviewed in Arentoft et al., (2004).

We believe that the high photometric precision of our survey was the key to
the above discovery for the upper main sequence stars. 
Without this high precision we would not have found any of these stars to be
variable. Clearly, the precision level of ones data sets the ultimate limit on
discovery of such low amplitude stars. Smith and Groote (2001) discovered Am stars in
Orion region clusters as young as 500 million year old while NGC 2301 is only
164 million years old. Thus, whether this $>$96\% variability fraction in the
upper main sequence is common in open cluster and field stars or specific to
this young, metal rich cluster awaits additional, high precision 
photometric surveys.

\subsubsection{Three New Variables with Extreme Colors}

Three stars caught our attention as being both extreme in color and variable.
The first, C0023, is a bright (B=14.8, R=11.5) red variable located very close
to the cluster center. Figure~\ref{c0023} shows our light curve for the star and a
zoomed version covering a single night. The light curve hints at being periodic
but with a period much longer than our dataset and reveals probable intra-night
variations of 0.02 mag over time scales of tens of minutes.

As a part of the SMARTS collaboration with Fred Walter (SUNY), we were able to
get red and blue spectra of C0023 in an attempt to discover its true nature.
The blue spectra were obtained on 30 Dec. 2004 UT and the red spectra on 06
Jan. 2005 UT. Figures~\ref{c0023b}\&\ref{c0023r} 
show the results. The blue region shows weak Ca II
H\&K and strong TiO absorption while the red spectrum is dominated by molecular
absorption as well with weak Na I and Ca I. Spectral identification was
performed via comparison to known types and C0023 was determined to be a M4III-M6III
star. Taking the B-R color, the R magnitude, and the assumption that C0023 is
a normal giant, it has an M$_V\sim$0 and therefore a distance
of $\sim$5 kpc. C0023 is therefore not a cluster member but lies well in
back of NGC 2301. We have already seen that the reddening to NGC 2301
is low (0.028 mag) but beyond the cluster much higher extinction probably exists.
This distant, very red and intrinsically bright M5III star no doubt suffers
from the post cluster extinction and we note, compared to normal M5III colors, 
that the star C0023
has an additional $\sim$0.4 mag of reddening in its B-R color.

Two other sources show extreme colors and are variable.
These are the two objects that fall near the white dwarf cooling track region
(see the CMD). Both sources are variable, but not (strictly) periodic, and both are blue in color.
C0758 is the fainter of the two at B=18.95, R=18.65 while the other, N0752, has
B=17.6 and R=18.1 but a large amplitude in its variability, showing an apparent 4
magnitude deep eclipse. Figure~\ref{wds} shows our R band light curves for 
these two sources. At this time, the true nature of these two sources and even 
if they are truly cluster members are both unknown.


\section{Amplitude and Period Statistics}

As a summary of our findings for this project, we present a set of histograms that
collect the information on color, period, and amplitude.
Figure~\ref{pvar} shows the distribution of our observed stars as a function 
of their B-R color.
We see that the peak in the color distribution falls near spectral type K0
(B-R=1.8) but stars from A to mid-M class are well represented.
The bottom panel of Figure~\ref{pvar} shows the fractional distribution 
for sources
we identify as variable. The histogram bin uncertainties are given by
$$ \sigma=\frac{N_{variable}}{N_{total}} \times \sqrt{\frac{1}{N_{variable}} + \frac{1}{N_{total}}} $$
and we see that the distribution yields a somewhat expected result. It is likely that 
blue stars are variable
mainly due to pulsations, red stars are variable mainly due to rotation and stellar
activity, with eclipsing binaries generally adding to the sample over 
most of the entire color range.
T05 found that the quartile variability of all the stars (without regard for
periodicity) follows a power law, with the cumulative fraction of stars with
variability less than $x$ being given by $f(<x) = 1 - 1.6~{\rm mmag} / x $ for $x>$1.6
mmag. 
A similar power law distribution was found for a sample of field 
stars by Everett et al., (2002). 
We remind the reader that Figure~\ref{pvar} does not
attempt to separate field stars and cluster members.

Figures~\ref{ampnp}\&\ref{ampp} present histograms of the amplitude 
(magnitude difference between the min/max points) 
of each variable star, separated into non-periodic and periodic bins. 
A few stars in both groups have amplitudes larger than 1 magnitude, but are not
plotted here to allow the lower amplitude majority to be effectively seen\footnote{
We calculate a value for the maximum
deviation as well as the mean weighted amplitude for each light curve 
.}. Figure~\ref{col_amp} presents the amplitudes observed in 
all non-periodic variables as a function of color. We see from
these three figures that both periodic and non-periodic variables 
show a peak in the
amplitude distribution at small values, $\la$0.1-0.15 magnitude and we note
that the bluest 
stars show a ``spike" in amplitude reaching up to about 1 magnitude. In general, most
variables are 
grumbly stars that contain the levels of variation that 
extra-solar planet transit hunters are interested in. This fact illustrates a major
hurdle to overcome when searching for small amplitude signals such as transits. 
It reinforces the need for observational searches of candidate transiting 
sources to extend over 
many tens of days or longer in order to allow any periodic signal (i.e., a
transit) to be separated from other similar amplitude noise.

Using our period finding method described above, Figure~\ref{per_hist}
presents the distribution of orbital period that we find in our stars that have been
deemed periodic. We see that shorter periods are three times more prevalent and the
distribution seems to flatten for period of 8 to 12 days. Due to our time sampling 
and total length of observation, our data set is only
sensitive to periods of $\sim$30 minutes to $\sim$12 days.
Some loss of sensitivity to longer periods is no doubt caused by our 
finite duration of
observation. 
However, it may be likely that shorter time scale variations are really 
more numerous as they will generally consist of lower amplitude intrinsic causes 
such as pulsation, starspots, etc., 
while longer periods usually consist of larger amplitude intrinsic 
(long period Miras, Cepheids, etc.) and/or
extrinsic (e.g., eclipses) causes. We do not distinguish in this figure 
the cause of the period but our data suggests that in general the shorter 
periods are mainly pulsational (bluer stars) and rotational while the longer periods are a 
combination of eclipsing stars and effects due to (slower) rotation or stellar activity.
To test this hypothesis, Figure~\ref{col_per} presents our determined periods as a function of
the stellar color. While not very definitive, we do see a grouping of blue stars which
have only short periods.

\section{Predictability of Variability}

Variability is a phenomenon that does not lend
itself easily to description by a single number or functional description. 
In T05, we presented our variability characterization in cumulative probabilities
via the statistical method of quartiles. Quartiles are much like medians as a
they provide a determination of a ``middle" value that is somewhat insensitive to
outliers. In fact, the second quartile equals the median value for a data set.
But in any statistical description of variability, statistical terms can become
confusing and even meaningless. For example, the term amplitude. 
This term is very ill defined in the presence
of noise and non-sinusoidal functions.

Often one uses the root mean square (rms) of a light curve (1$\sigma$) 
as a description of its level of variability, 
however, this term is dependent on outliers. For example,
how well does a 1$\sigma$ or 3$\sigma$ statistic describe a flare star which is
normally constant but shows an occasional flare. The time sampling of such an
event can change it from strictly having outliers and thus being
non-variable in some statistical tests to having a large amplitude deviation
and being a robust variable. Various other measures of location such as the mean
deviation (see Mighell and Burke, 1999) or the quartile deviation (which finds the
median of the middle 50\% of all values) have been tried with varying success.

We have chosen to present a statistical methodology with tries to 
overcome the leverage of outliers to make or break an object's variability (or
constancy) and to provide a robust measure of variability taking advantage of
our high precision measurements.
T05 found that the observed level of variability is well described by a power law
and they present an equation for the cumulative fraction at a specific variability
level (or higher) (see \S4).
Within the errors of T05, 
all sources are variable at 1.6 mmag (f=1), half are variable at x=3.2 mmag, etc.
These authors used quartile variabilities so we may wish to relate 
them to the more common rms or $\sigma$ levels of a light curve.  

The ``conversion" between quartile variability, rms, and amplitude strictly 
depends on the
distribution. If you have a random distribution, the relationship is roughly
that the quartile variability is 1.5 times smaller than the rms (1$\sigma$) for a 
Gaussian distribution and the
amplitude (for detection) may be 2-2.5 times larger 
($rms = \sim1.5 * \rm{quartile~variability}$ and $amplitude \sim2.5 * \rm{quartile~variability}.$)
The actual distribution of variability comes into play here
as one can get a variety of amplitudes if you
sample enough times. For a purely sinusoidal function, 
$amplitude = 1.4 * rms $ and $rms = \rm{quartile~variability}.$
These guidelines, however, say nothing about the nature of the 
variability: periodic,
multi-periodic, non-variables which are misclassified, objects such as flare
stars and eclipsing variables, or longer term variables such as carbon stars or
novae.
Using a statistical method with a preset cutoff (such as 2.5$\sigma$) is a formal 
method we will discuss below but keep in mind it is imperfect.

Taking T05's variability characterization at face value, how can we apply 
their result to studies
of  variability presented in the literature?  
Everett \& Howell (2001), for example, found that
17\% of the sources they observed (at their best precisions) 
were variable. They quote a best photometric
precision of $\sim$2-3 mmag and use a (normal distribution) variability 
criteria of $\sim$3$\sigma$ as a cutoff for conservative variability.
Making an assumption that on average over all amplitude types, 
it took at least 2.5*rms (7.5 mmag) to be possibly considered a variable
in the Everett \& Howell study, the T05 relation
would say that 78\% of all stars are variable at the $<$3$\sigma$
level concluding
that in the Everett \& Howell sample, 22\% sould be variable at their detection
level and above. This result is in good agreement with the Everett \& Howell 
observations.

We see that the predictive power of the T05 result (and in fact any predictor
of variability) depends not only on the level of
photometric precision reached in a survey, but on the level at which the survey
believes it 1) can detect true variability and 2) the criteria used as a cutoff
for sources to be considered variable. In this current work, our best
photometric precision is near 1-1.5 mmag, leading to a 
very conservative, statistically significant detection limit for variability 
at 3$\sigma$=4.5 mmag, or a quartile value of $\sim$3.0 mmag.
Our variability law predicts that at our detectability level 
we should find that 53.3\% of all our sources will be variable.
Our determination, presented in \S3.3, is that 56\% of all our sources are
variable, in good agreement with our observations.
Figures~\ref{ampnp}\&\ref{ampp} present our differential counts for
variability and we 
over-plot our predicted fit for variability as a function 
of amplitude. 
We see that our fits are a good
match to both non-periodic and periodic variables.

\section{Conclusion}

We have presented time series R band photometry of the young open cluster NCG2301. 
Our data consist of R band time series and single B measurements. We find that 56\% of all our
sources are variable and of these, 13\% are observed to be periodic based on a single pass
period search. 
All B-R color values seem to be have roughly the same percent of variability
with the majority of amplitudes
being 0.15 magnitudes or smaller. Of the periodic variables, shorter periods 
dominate and likely contain stars which are both intrinsic and extrinsic variables.
We find that essentially all of
the upper main sequence A and F stars are variable. 
These stars lie in the $\gamma$Dor and/or $\delta$Sct instability strips and
generally reveal multiple periodic structure. 
We have shown that we can robustly describe the fraction of variable sources, 
as a loose function of amplitude, expected to be present in a given photometric survey.
The cumulative fraction of variables in a survey is well fit by a power law
distribution.

There are many interesting single variables in our survey database which warrant 
additional
photometric and spectroscopic observations. 
All of the photometric data products are available from the first author. 
This includes the light curves, the B,R magnitudes, the star positions, 
and a complete description of the statistical tests applied to the time series
results. The present work shows the extreme power of many nights of
time allocated to a single object and the ability of high precision photometry to 
reveal new insights
about astronomical sources. The need to perform a similar observational program 
for other open
clusters with different ages, metal content, and Galactic location is clear.

\acknowledgements

The authors wish to thank Helmut Abt, Con Deliyannis, Sidney Wolff, and Bob Mathieu for
insightful discussions and drawing our attention to some specific helpful
references. The anonymous referee brought our attention to a number of details,
initially overlooked, which have been examined and have led to a better paper.
We also thank Fred Walter and the SMARTS consortium and Sebastien
Lepine for obtaining spectra for this paper.
The spectra of C0023 were obtained using the facilities of the SMARTS
consortium. Stony Brook University is a member of
the SMARTS consortium, which operates
the small telescopes at CTIO under contract to the NSF.
The data were obtained by SMARTS service observer Alberto Pasten.
Ken Mighell helped in a number of discussions about statistics.
We thank
the UH TAC for the generous allocation of telescope time. CV and RS had their
research
supported by the NOAO/KPNO Research Experiences for Undergraduates (REU)
Program which is funded by the National Science Foundation through Scientific
Program Order No. 3 (AST-0243875) of the Cooperative Agreement No. AST-0132798
between the Association of Universities for Research in Astronomy (AURA) and
the NSF.

\newpage

\begin{deluxetable}{ccc}
\tablenum{1}
\tablewidth{4.0in}
\tablecaption{Observed Point Sources in NGC 2301}
\tablehead{
 \colhead{Field}
 & \colhead{Designation}
 & \colhead{No. of Sources}
}
\startdata
\hline
Center &	C &	631  \\	
North &	N &	694  \\
South &	S  &      725 	 \\
Southeast &	SE &	660 \\ 	
East &	E &	678 \\
West &	W &	690 	 \\
Total & &	4078 \\
\hline
\enddata
\end{deluxetable}{}

\begin{deluxetable}{cccc}
\tablenum{2}
\tablewidth{2.5in}
\tablecaption{Some Constant Stars}
\tablehead{
 \colhead{Star}
 & \colhead{B}
 & \colhead{B-R}
 & \colhead{$\chi^2$/n}
}
\startdata
\hline
C0035  & 	13.7 &  1.4  &   0.88 \\	
C0102  & 	14.9 &  1.0  &   0.97 \\
C0246  & 	18.0 &  1.6  &   0.70 \\	
N0028  & 	15.2 &  1.4  &   0.99 \\	
N0094  & 	18.6 &  2.9  &   0.88 \\
N0272  & 	18.9 &  2.2  &   0.80 \\	
S0013  & 	14.0 &  0.8  &   1.10 \\	
SE0116  & 	17.3 &  1.5  &   0.69 \\				
SE0171  & 	18.7 &  2.5  &   0.76 \\	
E0040   &       15.2 &  1.2  &   1.99 \\	
E0471  &        21.0 &  2.4  &   1.46 \\
W0024  & 	15.2 &  1.3  &   0.91 \\ 	
W0046  & 	17.3 &  2.3  &   0.66 \\	
W0048  & 	15.1 &  1.0  &   0.51 \\
\hline
\enddata
\end{deluxetable}{}

\begin{deluxetable}{ccccc}
\tablenum{3}
\tablewidth{5.5in}
\tablecaption{Matched Region C Variable Stars}
\tablehead{
 \colhead{Kim et al.}
 & \colhead{Region C star}
 & \colhead{B-R}
 & \colhead{Kim et al. type}
 & \colhead{Phasing?}
}
\startdata
\hline
V1  & 	C0012 & 0.15 &	SPB star &	no \\ 	
V2  & 	C0004 & 0.52 &	Magnetic CP star? &	no  \\	
V3  & 	C0037 & 0.55 &	$\gamma$Dor star &	yes  \\	
V4  & 	C0042 & 0.52 &	$\gamma$Dor star &	no \\
V5  & 	- & -- &	Eclipsing Binary & -- \\
V6  & 	C0140 &	1.48 &	Eclipsing Binary &	yes \\ 
V7  & 	C0141 &	1.50 &	Eclipsing Binary &	yes  \\
V8  & 	C0163 &	1.78 &	Eclipsing Binary &	yes  \\
V9  & 	- & -- &	Eclipsing Binary & -- \\
\hline
\enddata
\end{deluxetable}{}

{}

\clearpage

\clearpage

\clearpage

\clearpage

%
 
\clearpage

\clearpage

\clearpage

\clearpage

\clearpage
\begin{figure}[t]
\epsscale{0.80}
\plotone{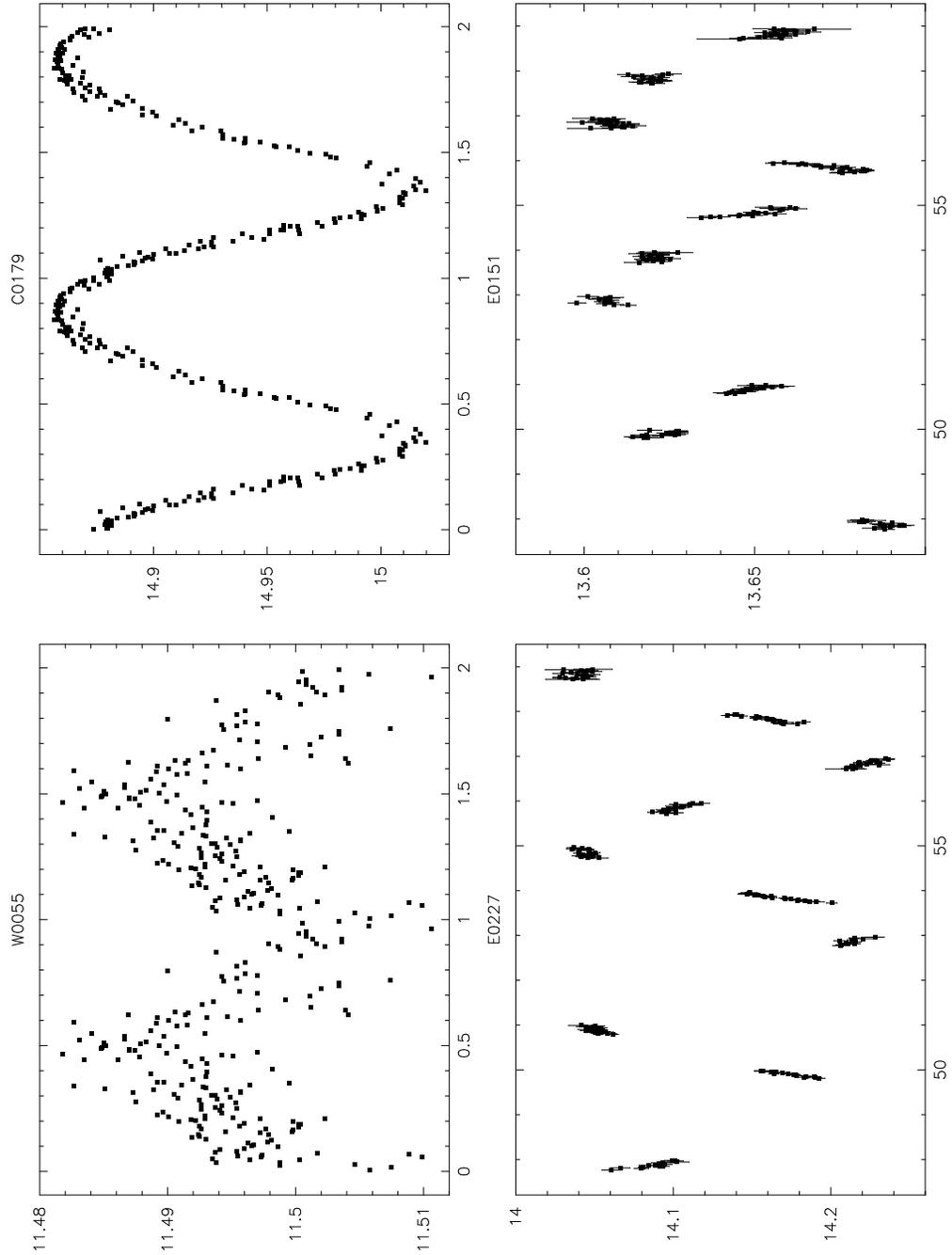}
\caption{
Four periodic variable
examples. The top two panels are phased on the best period determined (C0179 -- Period = 0.784
days, B=17.8, R=16.1; W0055 -- Period = 1.16 days, B=15.7, R=14.5) and no error bars are
plotted for clarity. The bottom two panels show unphased light curves for two longer period
likely eclipsing stars (E0227 -- Period = 3.2 days, B=18.7, R=17.0; E0151 -- Period = 3.88
days, B=18.6, R=16.6). The y-axis values are instrumental magnitude and the x-axis are phase
(top) and HJD-2453000 days (bottom).
\label{pervar}}
\end{figure}

\clearpage
\begin{figure}[t]
\epsscale{0.80}
\plotone{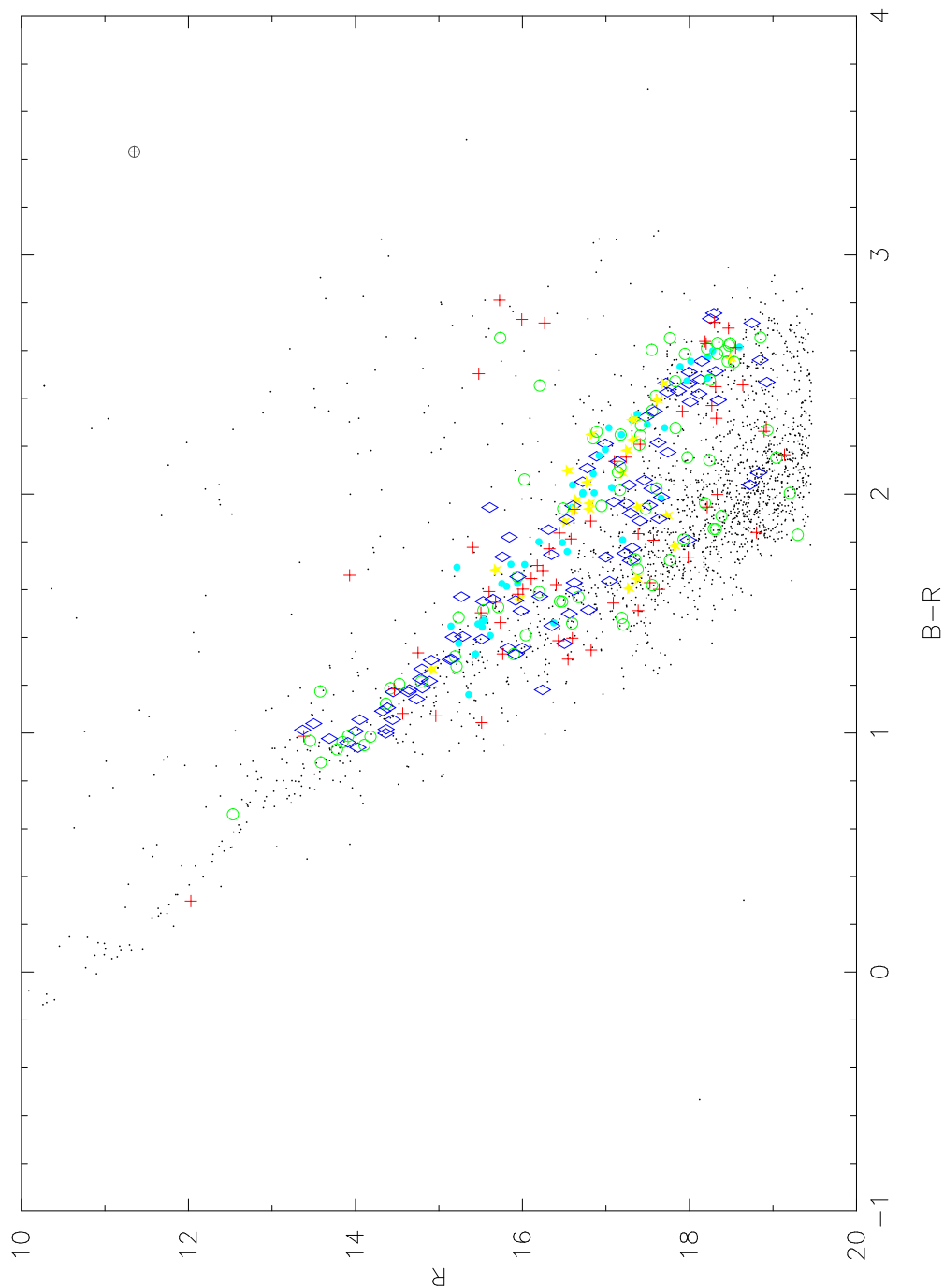}
\caption{
Periodic stars in NGC 2301. 
Sources that are constant or variable (but not periodic) are shown as
small black dots. Periodic sources are shown as colored symbols according 
to the
following scheme: Circle with cross (see Fig. 12);
red cross = period $<$1 day;
green circle = period $>$1 day but $<$3 days;
blue diamond = period $>$3 days but $<$7 days;
filled cyan dot = period $>$7 days but $<$10 days;
yellow star = period $>$10 days;
The lack of more periodic stars in the upper main sequence is likely a 
function of our simple one-pass period search and the fact
that these pulsating stars are multi-periodic.
The lower main sequence is nicely outlined with periodic 
variables and we see fewer periodic variables in the non-cluster
member ``wedge'. 
\label{cmdper1}}
\end{figure}

\addtocounter{figure}{-1}
\clearpage
\begin{figure}[t]
\epsscale{0.80}
\plotone{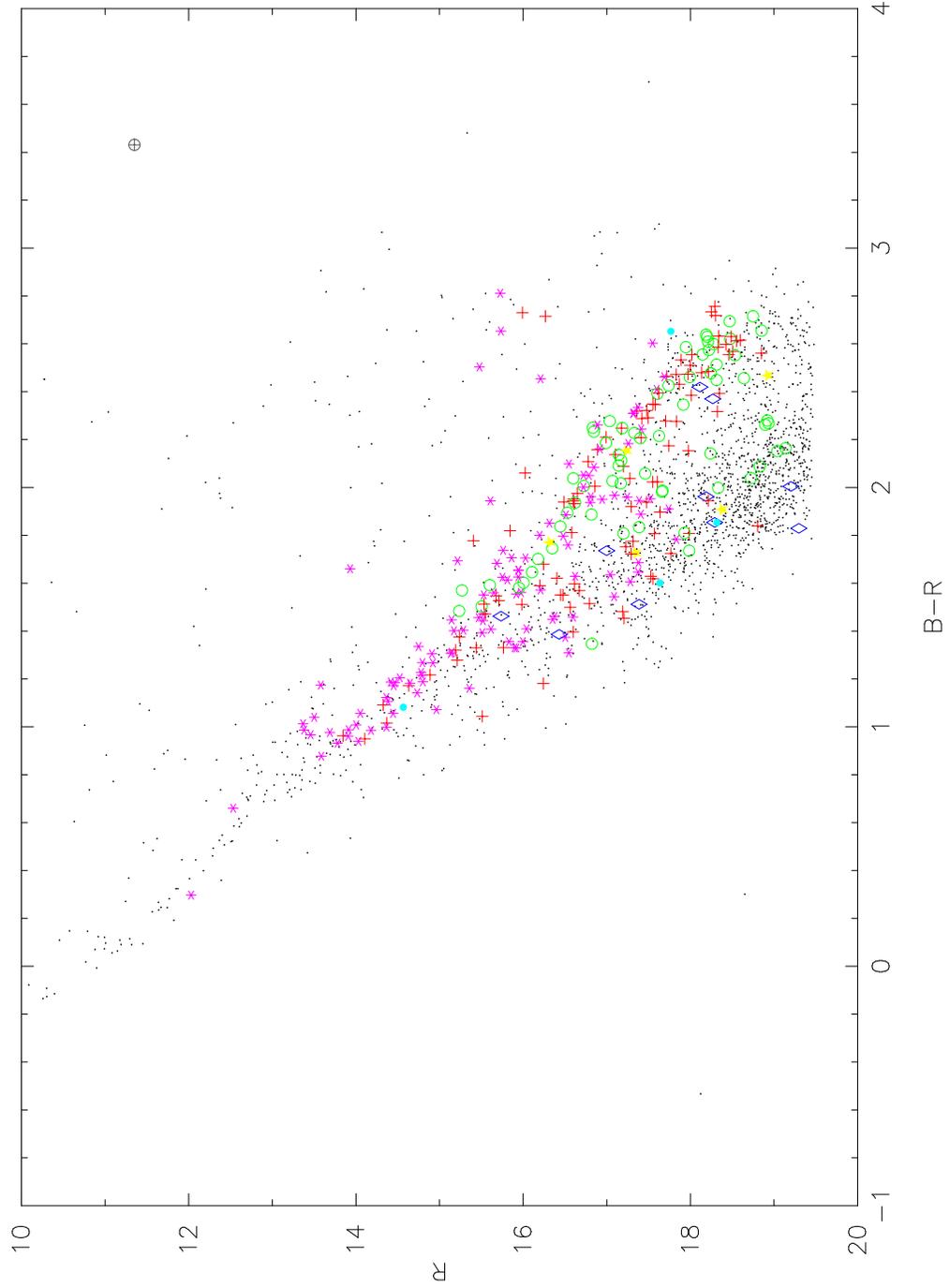}
\caption{
Periodic stars in NGC 2301 sorted by amplitude. 
Sources that are constant or variable (but not periodic) are shown as
small black dots. The amplitude color scheme is the same as in Figure 5.
\label{cmdper2}}
\end{figure}

%
 
\clearpage
\begin{figure}[t]
\epsscale{0.80}
\plotone{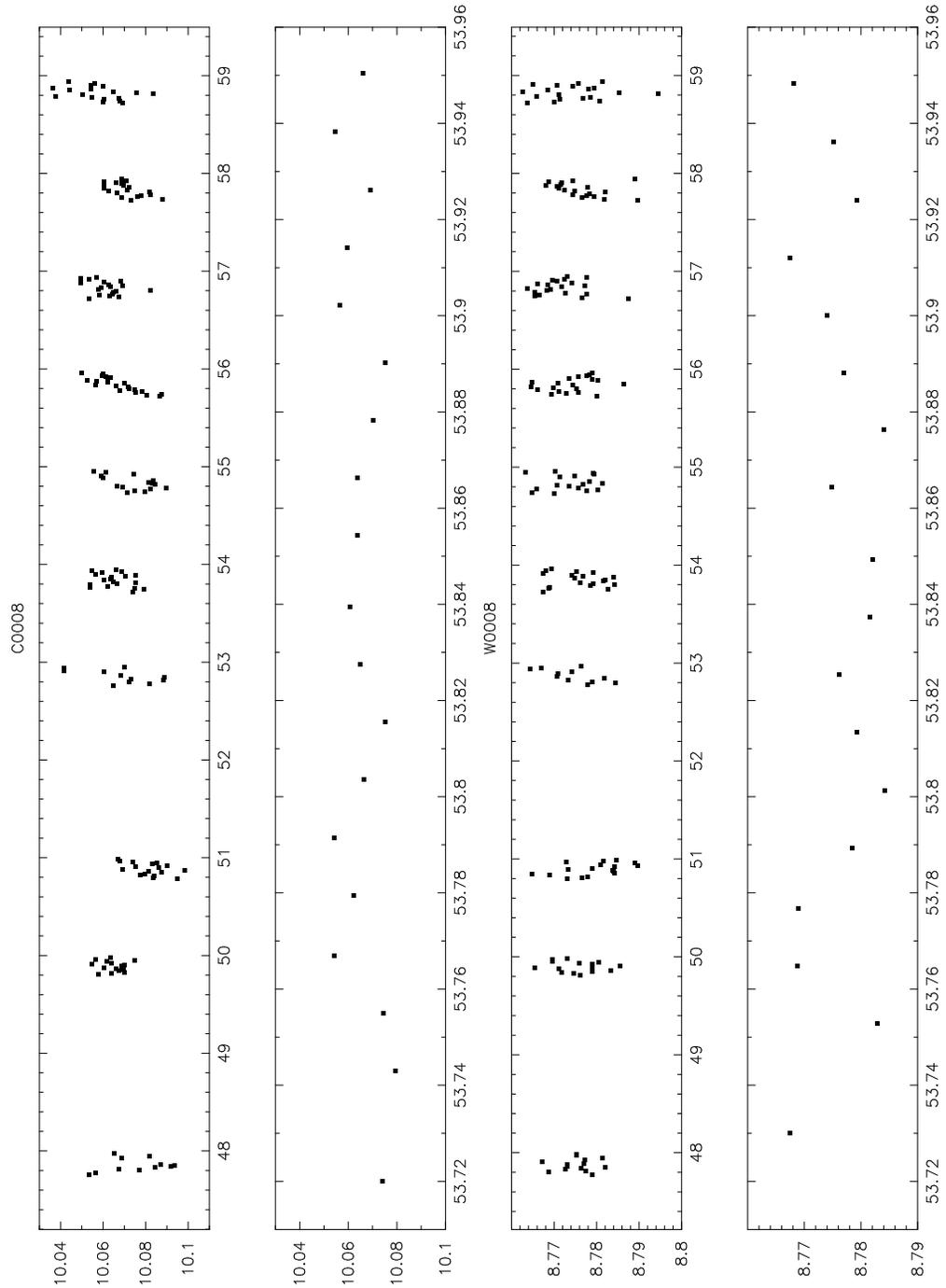}
\caption{
Two examples of
multi-periodic upper main sequence stars C0008 and W0008. The top two panels show
the star C0008 as the full light curve with a single night of data plotted below.
The bottom two panels are likewise but for the star W0008. C0008 (W0008) has a 
best fit single
period of 0.488 (0.041) days and an R magnitude of 11.2 (11.8) 
with both B-R values near 0. 
The upper main sequence
stars in NGC 2301 lie in the $\delta$ Sct and/or $\gamma$ Dor instability strips.
The
y-axis is instrumental magnitude and the x-axis is HJD-2453000 days. 
\label{mulper}}
\end{figure}

\clearpage
\begin{figure}[t]
\epsscale{0.80}
\plotone{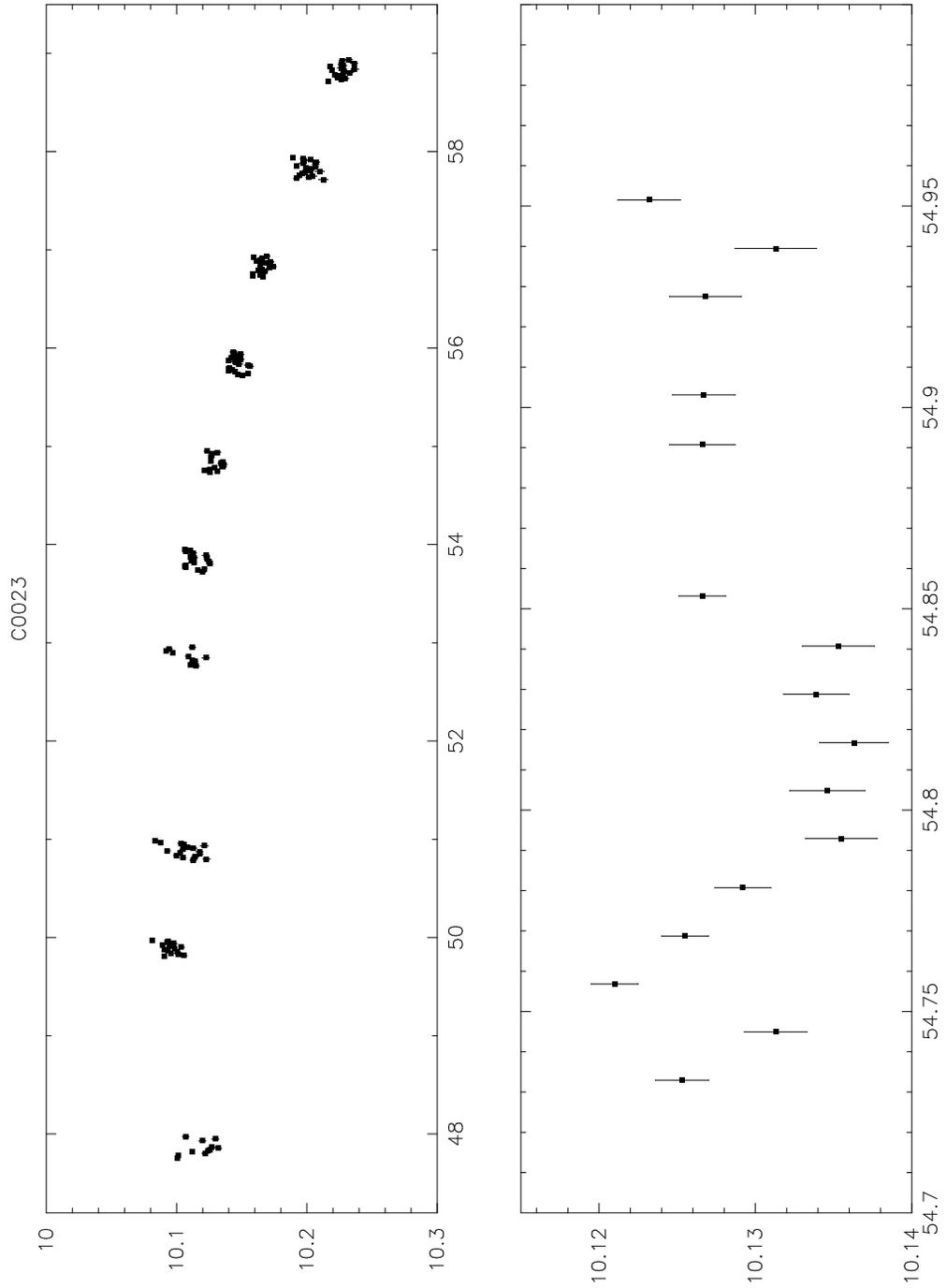}
\caption{
Light curve of the suspected Mira
variable C0023. Both panels show data for this star with the lower panel showing a single
night. Note that each night appears to have short term variations in addition to the 
longer term sine-like
trend. The y-axis is instrumental magnitude and the x-axis is HJD-2453000 days. 
\label{c0023}}
\end{figure}

\clearpage
\begin{figure}[t]
\epsscale{0.80}
\plotone{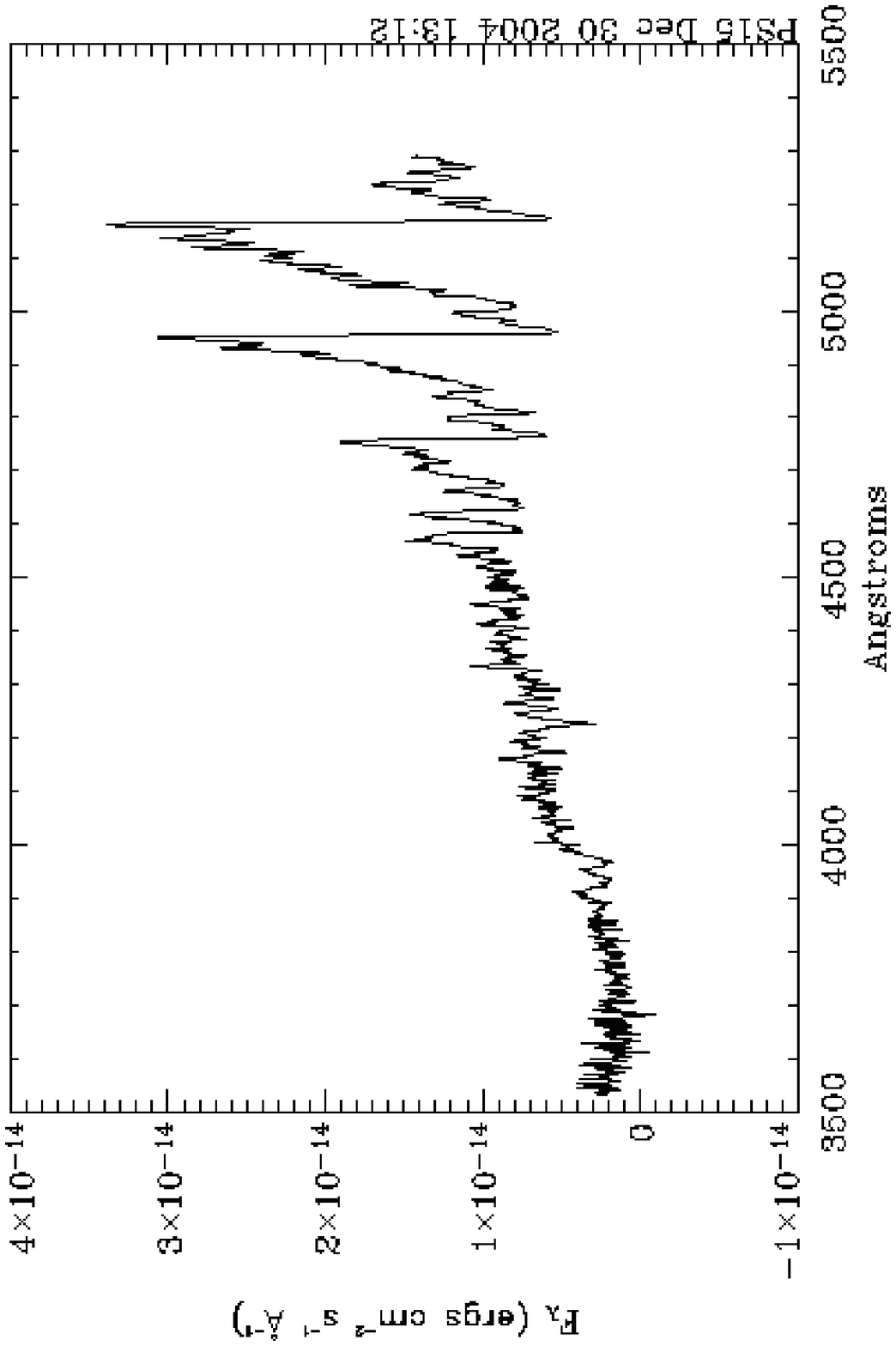}
\caption{
Blue spectrum of C0023.
Weak Ca II H\&K and strong TiO absorption are seen. This spectrum is consistent 
with an M4III star 
\label{c0023b}}
\end{figure}

\clearpage
\begin{figure}[t]
\epsscale{0.80}
\plotone{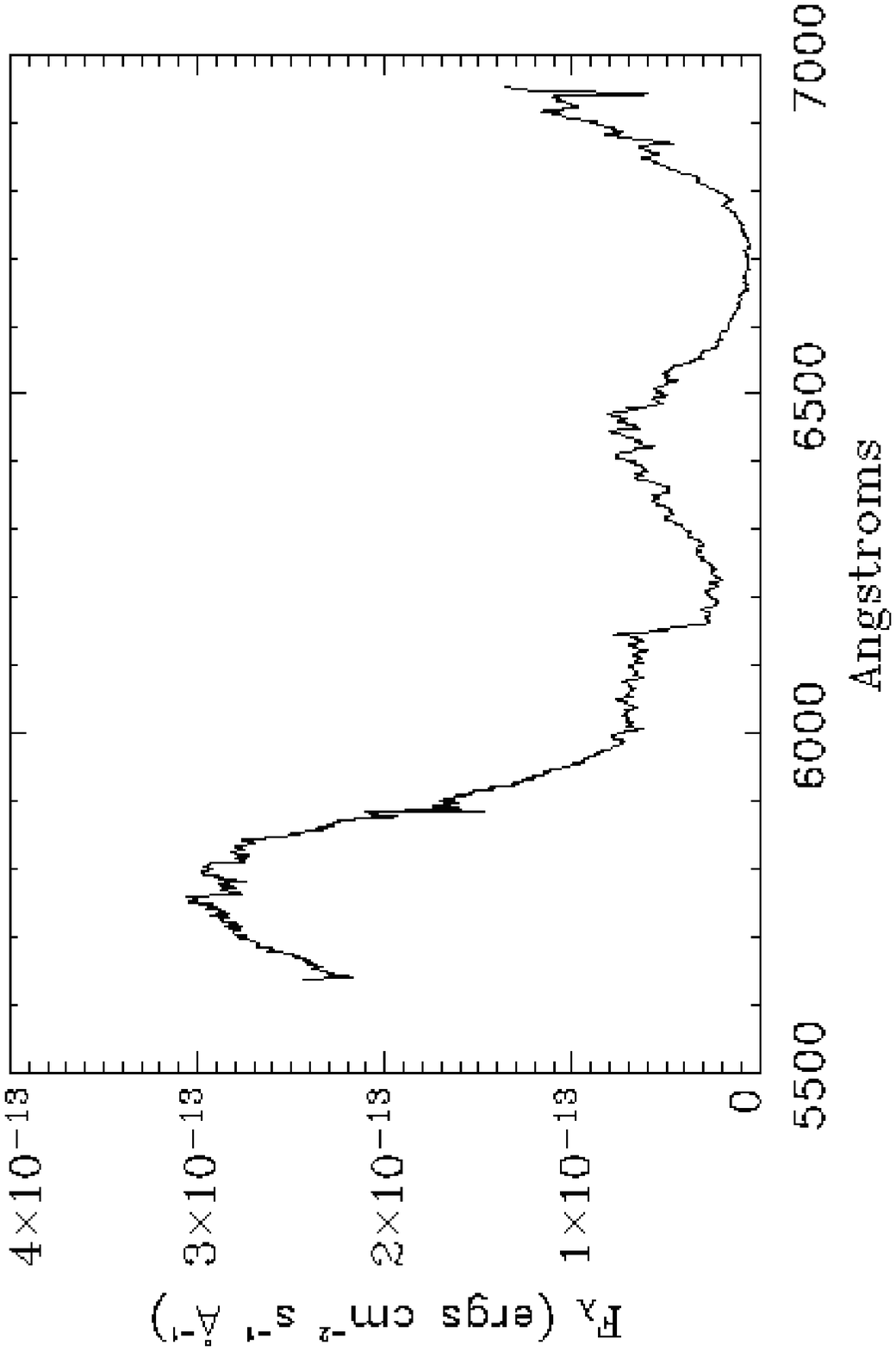}
\caption{
Red spectrum of C0023.
Weak absorption of Na I and Ca I are seen as well as strong molecular absorption. 
This spectrum is consistent with an M6III star.
\label{c0023r}}
\end{figure}

\clearpage
\begin{figure}[t]
\epsscale{0.80}
\plotone{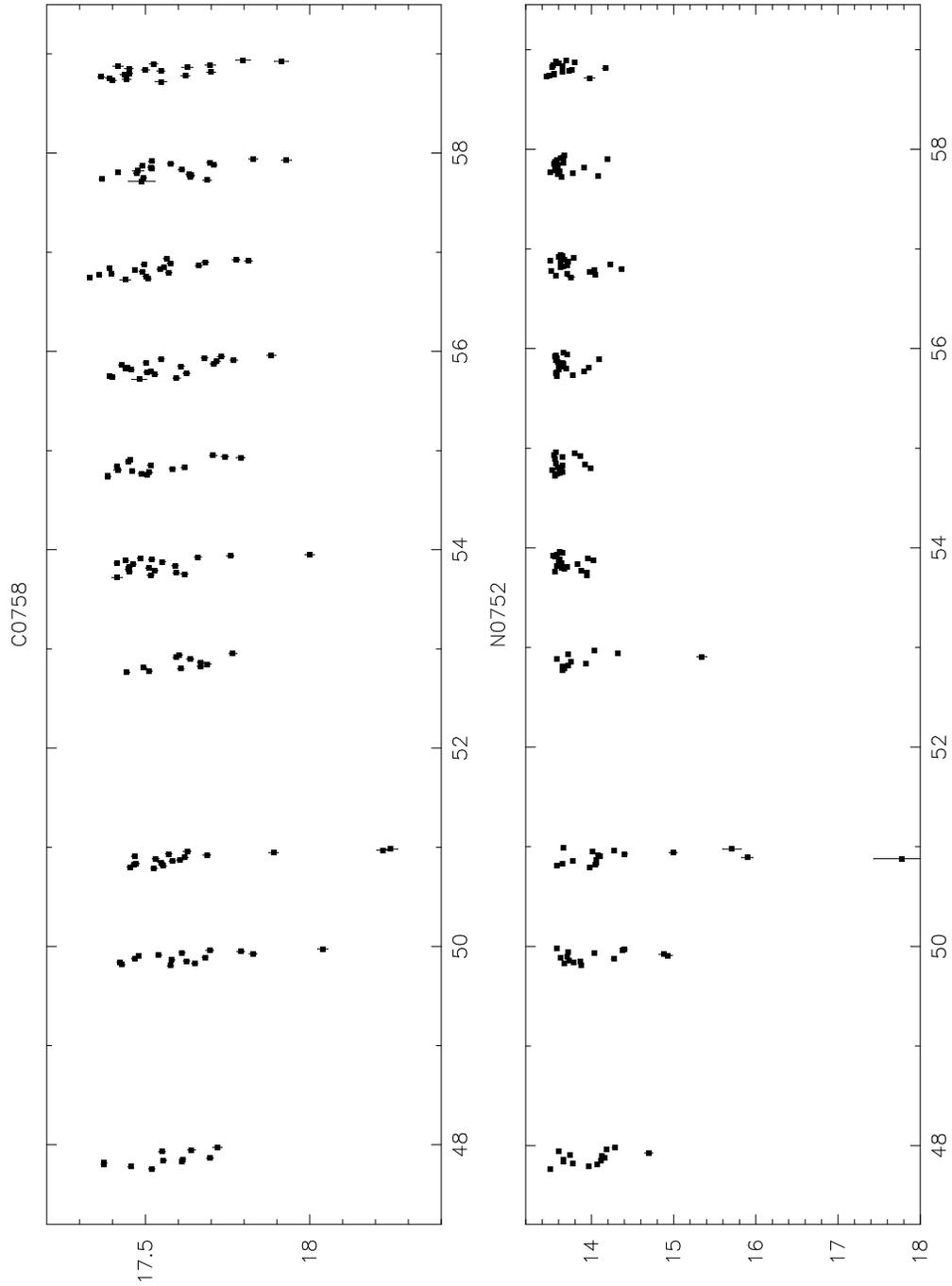}
\caption{
Light curves for the two
white dwarf candidates,
C0758 and N0752. Both stars seem to be highly modulated and show evidence for a possible
eclipse, N0752 in particular.
If these are eclipsing systems, the period appears to be fairly short, $\la $1 day.
The y-axis is instrumental magnitude and the x-axis is HJD-2453000 days.
\label{wds}}
\end{figure}

\clearpage
\begin{figure}[t]
\epsscale{1.00}
\plotone{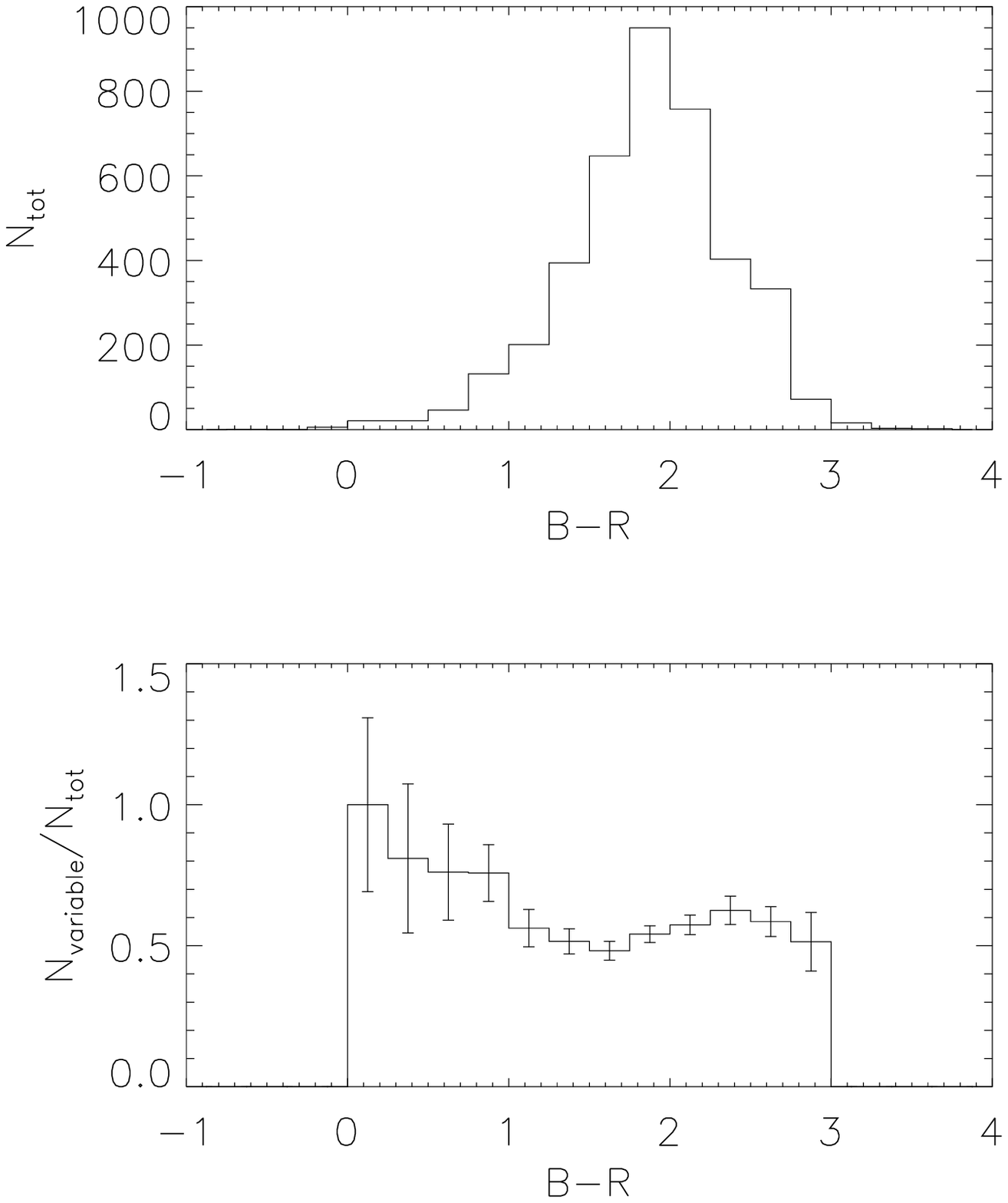}
\caption{
Histogram of all our program stars by number (top) and by 
fraction that are variable (bottom) as a
function of their B-R color. The distribution peaks near K0V in total 
number and
the percent variability is relatively flat across all colors with possible 
slight increases for blue
(mainly pulsating) and red (mainly rotation and stellar activity) stars.
\label{pvar}}
\end{figure}

\clearpage
\begin{figure}[t]
\epsscale{1.00}
\plotone{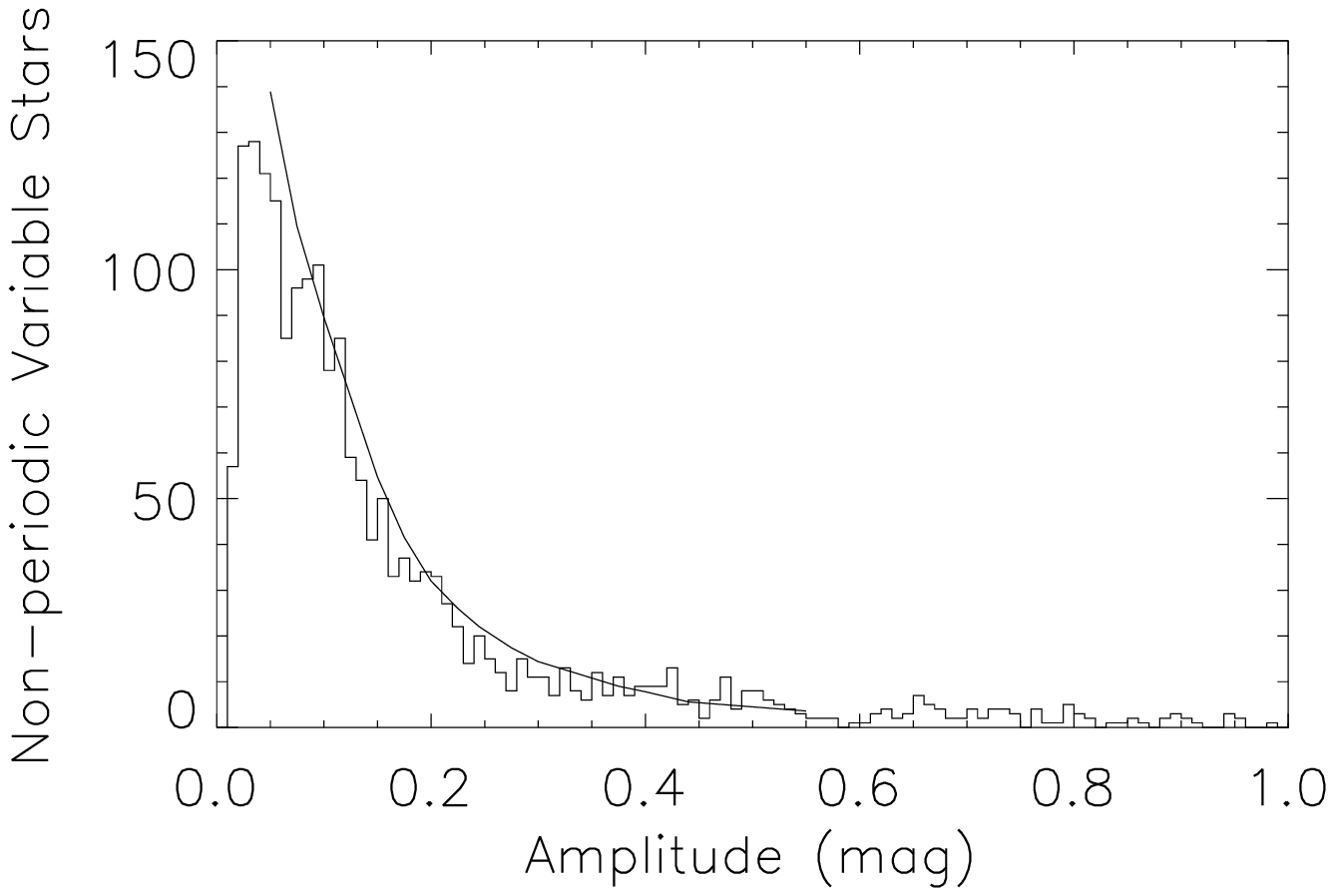}
\caption{
Histogram of the non-periodic variable stars showing the
amplitude of their light curve. We note a rapid fall off at larger
amplitude and a peak in the variations for amplitudes of $\la$0.15 magnitude.
The solid line is a fit to the distribution based on our prediction equation as
discussed in \S4.
\label{ampnp}}
\end{figure}

\clearpage
\begin{figure}[t]
\epsscale{1.00}
\plotone{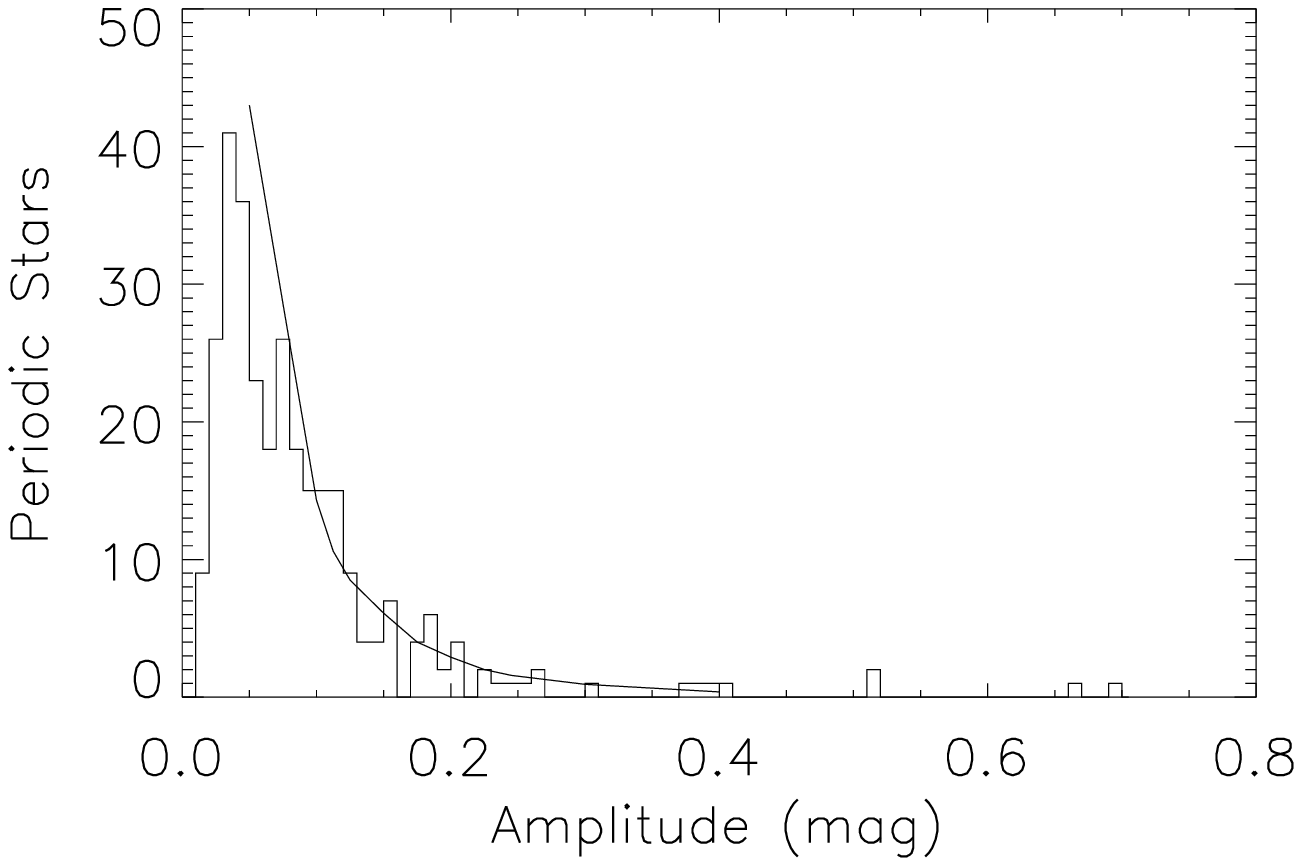}
\caption{
Histogram of the periodic variable stars showing the
amplitude of their light curve. We note a rapid fall off at larger
amplitude and a peak in the variations of $\la$0.1 magnitude.
The solid line is a fit to the distribution based on our prediction equation as
discussed in \S4.
\label{ampp}}
\end{figure}

\clearpage
\begin{figure}[t]
\epsscale{1.00}
\plotone{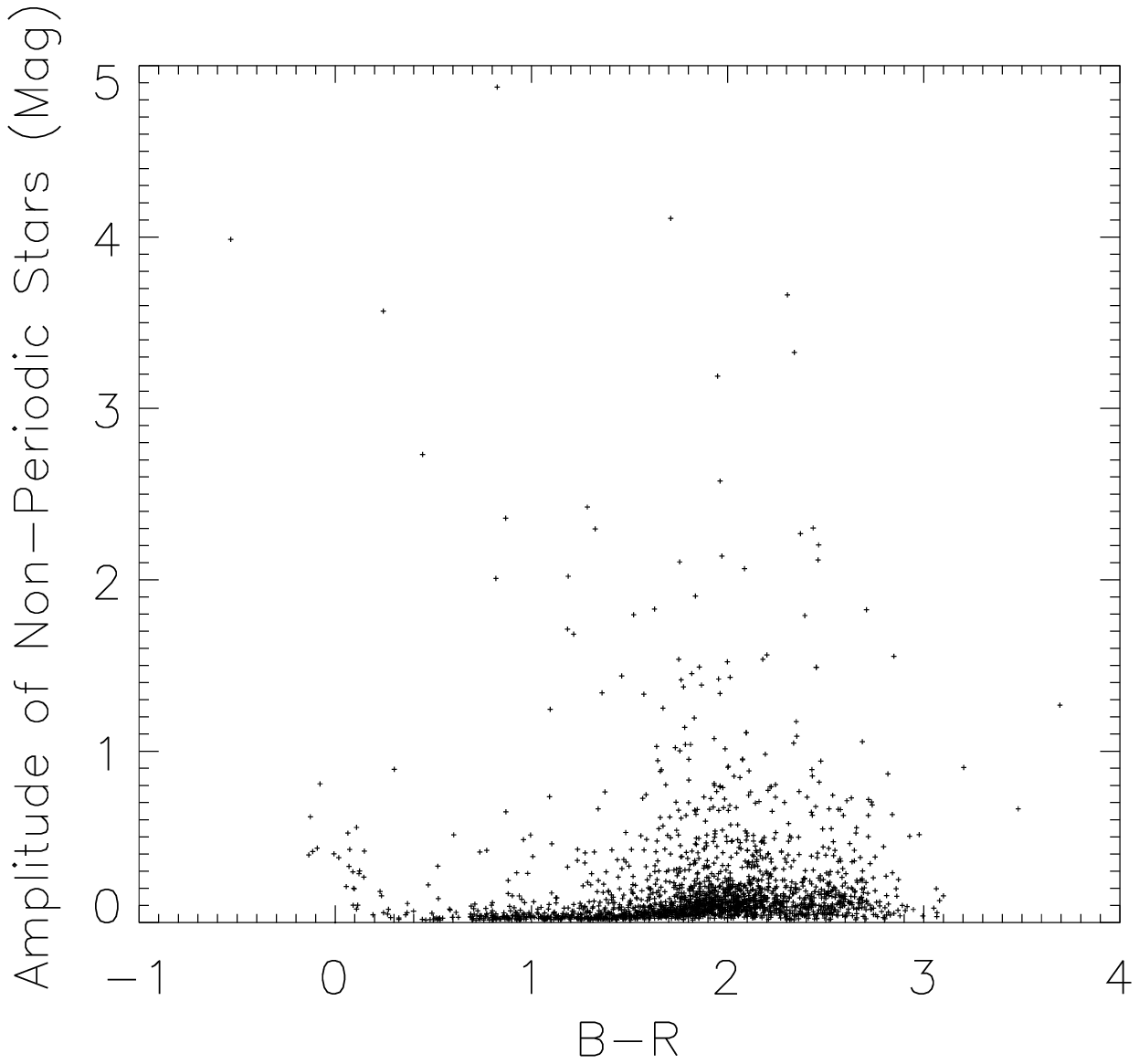}
\caption{
B-R color for our sample of non-periodic variables in all six fields. 
The majority of sources show low level (few tenths of a  magnitude) variations
with a broad range of color for the larger amplitude objects. Note the
small blue spike near B-R=0 showing amplitudes up to about one magnitude. Some of these
sources are multi-periodic pulsating stars on the upper main sequence.
\label{col_amp}}
\end{figure}

\clearpage
\begin{figure}[t]
\epsscale{0.80}
\plotone{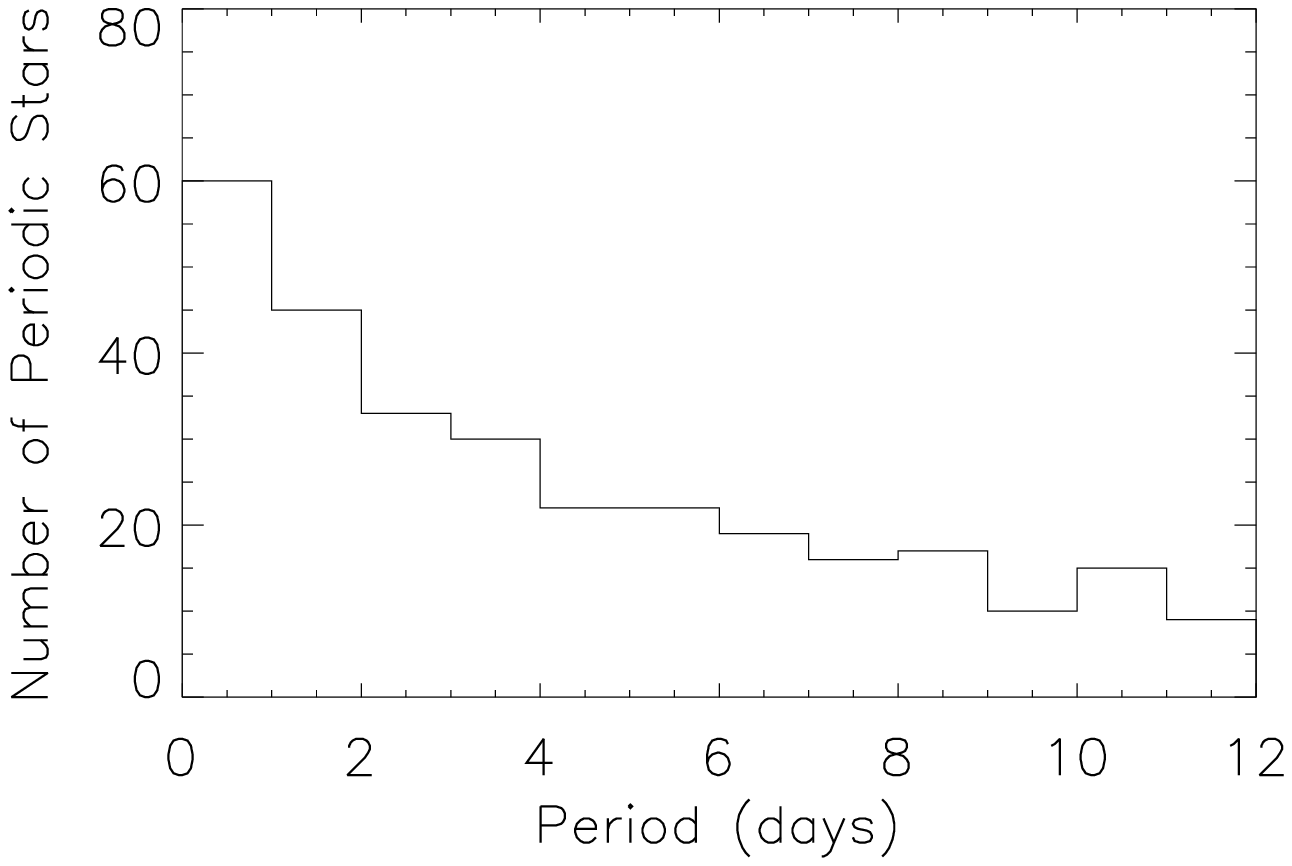}
\caption{
Histogram of the orbital period for the periodic variables in our sample.
We note a smooth and steady decline from short to long orbital 
period (see text).
\label{per_hist}}
\end{figure}

\clearpage
\begin{figure}[t]
\epsscale{1.00}
\plotone{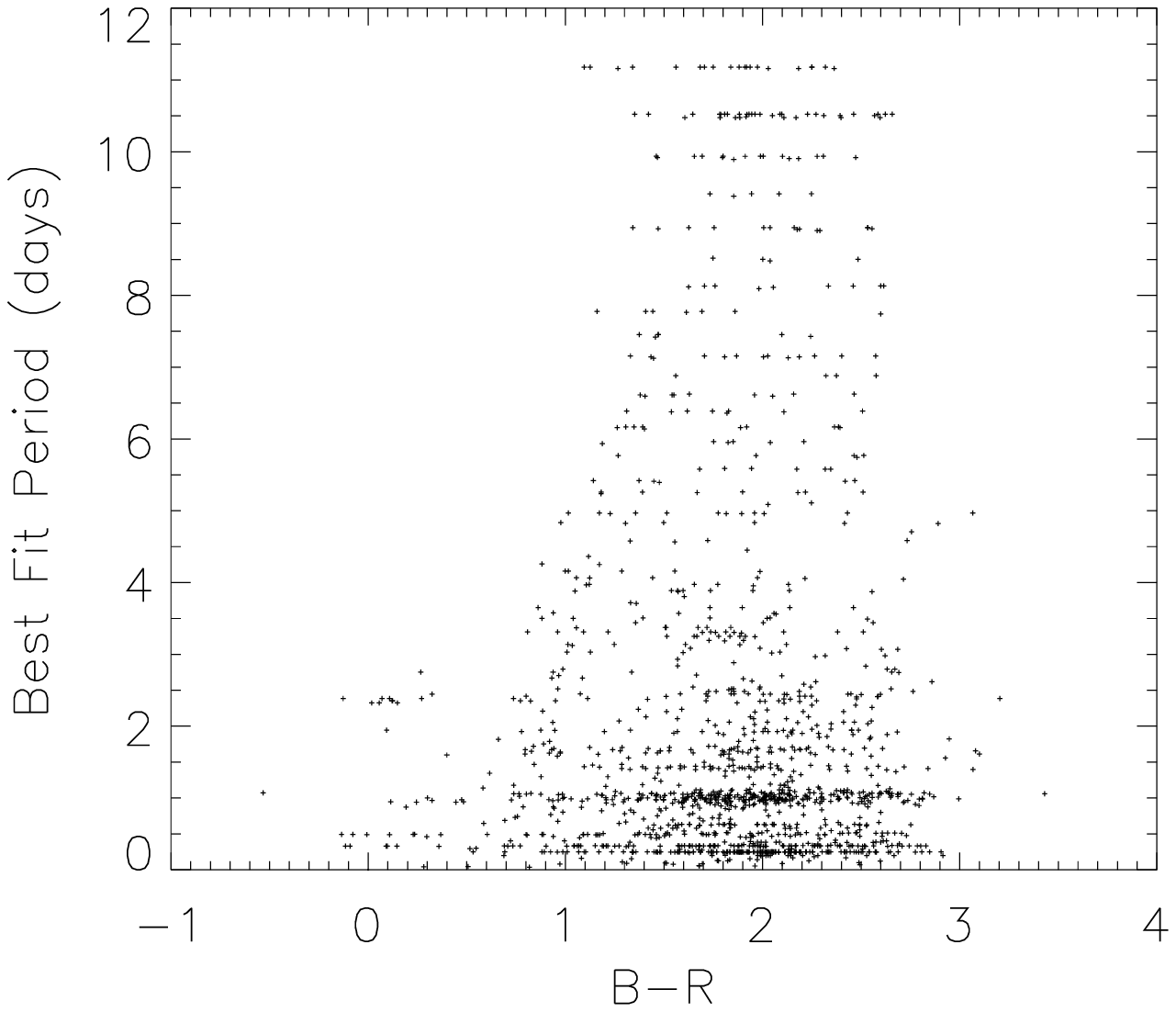}
\caption{
Our periodic variable sources as a function of color. The entire range of color is well
represented with a small spike for the bluest stars. At the longer periods (up
to 3 days) it may be that these blue stars represent binary systems on the 
upper main sequence.
The quantum nature of the period distribution is an artifact of our single pass, 
gross period search process.
\label{col_per}}
\end{figure}

\end{document}